\documentclass[twocolumn]{aastex631} 
\usepackage[utf8]{inputenc}
\usepackage{amssymb}
\usepackage{graphics}

\graphicspath{{./}{figs/}}

\accepted{May 26, 2025}
\submitjournal{ApJ}

\shorttitle{SLAMS at Astrophysical Shocks}
\shortauthors{Zekovi\'c et al.}

\begin{document}

\title{SLAMS-Propelled Electron Acceleration at High-Mach Number Astrophysical Shocks}

\correspondingauthor{Vladimir Zekovi{\'c}}
\email{vladimir.zekovic@matf.bg.ac.rs}

\author[0000-0002-4985-3253]{Vladimir Zekovi{\'c}}
\affiliation{Department of Astronomy, Faculty of Mathematics, University of Belgrade \\     Studentski trg 16, 11000 Belgrade, Serbia}
\affiliation{Department of Astrophysical Sciences, Princeton University \\
    Princeton, NJ 08544, USA}

\author{Anatoly Spitkovsky}
\affiliation{Department of Astrophysical Sciences, Princeton University \\
    Princeton, NJ 08544, USA}
    
\author{Zachary Hemler}
\affiliation{Department of Astrophysical Sciences, Princeton University \\
    Princeton, NJ 08544, USA}

\begin{abstract}
    Short Large Amplitude Magnetic Structures (SLAMS) are frequently detected during spacecraft crossings over the Earth bow shock. We investigate the existence of such structures at astrophysical shocks, where they could result from the steepening of cosmic-ray (CR) driven waves. Using kinetic particle-in-cell simulations, we study the growth of SLAMS and the appearance of associated transient shocks in the upstream region of parallel, non-relativistic, high-Mach number collisionless shocks. We find that high-energy CRs significantly enhance the transverse magnetic field within SLAMS, producing highly inclined field lines. As SLAMS are advected towards the shock, these fields lines form an intermittent superluminal configuration which traps magnetized electrons at fast shocks. Due to their oscillatory nature, SLAMS are periodically separated by subluminal gaps with lower transverse magnetic field strength. In these regions, electrons diffuse and accelerate by bouncing between the shock and the approaching SLAMS region through a mechanism that we call quasi-periodic shock acceleration (QSA). We analytically derive the distribution of electrons accelerated via QSA, $f(p)\sim p^{[-4.7,-5.7]}$, which agrees well with the simulation spectra. We find that the electron power law remains steep until the end of our longest runs, providing a possible explanation for the steep electron spectra observed at least up to GeV energies in young and fast supernova remnants.
\end{abstract}


\section{Introduction}

Astrophysical shocks driven by high-energy explosions are thought to be fast, having Mach numbers of $\gtrsim 100$. However, our understanding of shock evolution and electron acceleration is based on the results of particle-in-cell (PIC) simulations of shocks with Mach numbers of typically $< 40$. In such scenarios (e.g., \citealt{PEA}), the cosmic-ray-driven waves (\citealt{CR_streaming_inst}) saturate at lower levels of field amplification, $B_\perp \sim 2-3\,B_0$, where $B_\perp$ and $B_0$ are the strengths of the perpendicular (to the shock normal) field and the base field (which can be inclined relative to the shock normal), respectively. Hybrid shock simulations\footnote{In hybrid method, electrons are considered as massless fluid and ions as particles, which allows a much longer shock evolution compared to PIC, but at the cost of losing the physics of electron acceleration.} show that the amplification increases with the Mach number, reaching a factor of $\sim10$ in the case of $M\sim 100$ (\citealt{DamAnatoly2014}). These non-linear waves have a quasi-periodic field structure, which later forms density cavities via the filamentation (\citealt{filamentation_inst}, \citealt{cavities2013}) or cavitation instability \citep{cavitation_inst}. Similar non-linear structures preceded by the non-resonant modes appear in magnetohydrodynamic (MHD) simulations (\citealt{bell_inst}), where a prescribed current of high-energy cosmic rays (CRs) amplifies these structures to even higher levels ($\sim 100$). In the case of fast shocks found in young supernova remnants (SNRs), the amplified field within non-linear structures can intermittently become superluminal for adiabatic\footnote{Adiabatic (or magnetized) electrons are confined to the magnetic field. Their Larmor radius is small compared to the wavelength of the driven modes or the size of SLAMS.} (magnetized) electrons, meaning that electrons confined to the magnetic field would have to move faster than the speed of light along the highly inclined field lines to avoid being advected downstream. At very fast (usually relativistic) oblique shocks where the ambient field inclination creates a persistent superluminal configuration, electrons are generally limited to few cycles of shock drift acceleration (SDA) and, therefore, cannot reach CR energies (\citealt{rel_mag_shocks}). At non-relativistic high-Mach number oblique shocks, electrons can get energized or accelerated by  a variety of mechanisms. Shock reflected electrons can drive the waves on their own scales in the foreshock region. Electrostatic waves which are the fastest growing, enable electrons ``surfing'' on the leading shock edge to accelerate by the mechanism of shock-surfing acceleration \citep{ssa}. The firehose and whistler modes which are further driven by reflected electrons lead to electron scattering and return to the shock (\citealt{MarioAnatoly2011}, \citealt{qperp_1d_acc}). The mechanism of electron acceleration on such waves is known as the stochastic shock drift acceleration (SSDA; \citealt{SSDA}). The results of recent kinetic simulations (\citealt{Matsumoto}, \citealt{Amano}) showed that SSDA significantly contributes to electron acceleration at high-Mach number quasi-perpendicular shocks. These mechanisms lead to the generation of a power-law electron spectra with steep slopes. However, the maximum attainable energy in such acceleration processes is limited by the scale of the waves.

On the other hand, satellite measurements show that quasi-periodic, non-linear field structures are quite common in the upstream of the Earth's bow shock. They are known as Short Large-Amplitude Magnetic Structures (SLAMS) and represent a strongly non-linear phenomenon manifested by periodic enhancements of the magnetic field in front of quasi-parallel shocks (\citealt{SLAMS_init}). SLAMS are frequently detected by spacecraft in the solar wind (\citealt{SLAMS_obs_init}, \citealt{SLAMS_Wilson}), where they appear as field-amplitude pulsations on the scale of proton Larmor radius. SLAMS grow as a result of non-linear steepening of right-handed waves driven by ions reflected from the shock (\citealt{ionion_inst}). Faster and stronger SLAMS are even capable of triggering transient shocks in the solar wind. Previously, SLAMS amplified up to $\sim5-6\,B_0$ have been studied at low-Mach number shocks ($M_{\rm A} \sim 5$) with 2D kinetic simulations, both in the hybrid approach (\citealt{SLAMS_2dhybrid}) and with full PIC simulations (\citealt{SLAMS_PIC}). In contrast to the Earth bow shock where SLAMS are driven by reflected ions, we expect the SLAMS at fast, high-Mach number astrophysical shocks to be driven by accelerated CRs. Since CRs are much more energetic than reflected ions, even larger field amplification is expected in SLAMS at astrophysical shocks. Such structures induce highly inclined magnetic field lines which tend to become superluminal at fast shocks and thus significantly increase the energy threshold for electron injection into diffusive shock acceleration (DSA; \citealt{dsa_axford}, \citealt{dsa_bell}, \citealt{dsa_bland_ostr}).
Therefore, it is important to investigate the nature of such SLAMS and to determine their role in particle acceleration at quasi-parallel, high-Mach number ($\gtrsim 80$) astrophysical shocks. 

In this paper, we use the fully kinetic PIC method to study the formation of SLAMS and self-consistent electron acceleration in high-Mach number shocks. Our results from a large set of 1D and 2D runs with different shock parameters indicate that SLAMS grow and evolve in the high-Mach number regime, where they significantly alter the electron acceleration and diffusion processes. In Sec.~\ref{sec:setup} we give an overview of our shock simulations and then discuss the properties of SLAMS that appear in our runs in Sec.~\ref{sec:slams}. In Sec.~\ref{sec:elacc} we show the simulation spectra and introduce a novel mechanism for electron acceleration by SLAMS, which we call \textit{quasi-periodic shock acceleration}. Finally, in Sec.~\ref{sec:sum} we discuss the implications of this mechanism for power-law slopes of radio-synchrotron spectra in young SNRs.

\section{Simulation Setup}\label{sec:setup}

\begin{deluxetable*}{c c r c r r r l l c}
    \tabletypesize{\scriptsize}
    \tablewidth{\textwidth} 
    \tablecaption{\label{tab1} The parameters shown for each simulation run are: run number, ion-to-electron mass ratio, Alfv\'en Mach number, shock velocity in the laboratory frame, electron magnetization (the ratio of magnetic to kinetic energy density), simulation end time, total number of particles per cell, size of the simulation domain at the end of simulation, type of SLAMS in the upstream, and amplification of the transverse field, respectively. Runs 1--6 are in 1D, and Runs 7 and 8 are in 2D geometry.}
    \tablehead{
        \colhead{Run} & \colhead{$m_i/m_e$} & \colhead{$M_{\rm A}$} & \colhead{$v_{\mathrm{sh}}$ [c]} & \colhead{$\sigma_e$} & \colhead{$T_{\rm end} [\omega_{pe}^{-1} |\,\omega_{ci}^{-1}]$} & \colhead{$N_\mathrm{ppc}$} & \colhead{$N_x \times N_y \mathrm{[cells]}$} & \colhead{type of SLAMS} & \colhead{$B_\perp / B_0$ (amplification)}
        }
    \startdata 
    1 & 32 & 80 & 0.133 & $8.9\cdot 10^{-5}$ & $7.37\times 10^5\,|\,217$ & 1024 & $7.1 \cdot 10^5 \times 2$ & developed (persistent) & $\gtrsim$10 \\
    2 & 32 & 40 & 0.067 & $8.9\cdot 10^{-5}$ & $5.36\times 10^5\,|\,158$ & 1024 & $3.4 \cdot 10^5 \times 2$ & developed & $\sim 7$ \\
    3 & 32 & 20 & 0.033 & $8.9\cdot 10^{-5}$ & $7.51\times 10^5\,|\,221$ & 1024 & $2.8 \cdot 10^5 \times 2$ & weak & $3-5$ \\
    4 & 100 & 80 & 0.133 & $2.8\cdot 10^{-4}$ & $4.91\times 10^5\,|\,82\,\ $ & 1024 & $7.6 \cdot 10^5 \times 2$ & developed & $\gtrsim$10 \\
    5 & 32 & 200 & 0.33 & $8.9\cdot 10^{-5}$ & $1.23\times 10^5\,|\,36\,\ $ & 1024 & $4.3 \cdot 10^5 \times 2$ & initial (transient) & $\gtrsim$20 \\
    6 & 32 & 300 & 0.267 & $2.5\cdot 10^{-5}$ & $1.50\times 10^6\,|\,234$ & 1024 & $8.1 \cdot 10^5 \times 2$ & developed (persistent) & $\gtrsim$30 \\
    \hline
    7 & 32 & 80 & 0.267 & $3.6\cdot 10^{-4}$ & $1.03\times 10^5\,|\,61\,\ $ & 32 & $8.9 \cdot 10^4 \times 11200$ & evolving (with cavities) & $\gtrsim$10  \\
    8 & 32 & 80 & 0.267 & $3.6\cdot 10^{-4}$ & $3.47\times 10^5\,|\,204$ & 32 & $1.1 \cdot 10^5 \times 2800$ & developed (with cavities) & $\gtrsim$10  \\
    \hline
    \enddata
    \tablecomments{In all runs we set the sonic Mach number $M_{\rm S} = v_{\mathrm{sh}} / c_{\rm S}$ $\left( \mathrm{with}\ c_{\rm S} = \sqrt{\frac{5}{3} \, \frac{k (T_i + T_e)}{m_i} } \right)$ to be equal to the Alfv\`en Mach number $M_{\rm A} = v_{\mathrm{sh}} / v_{\rm A} = \beta_{sh} \sqrt{\frac{m_i}{m_e}} (\sqrt{\sigma_e})^{-1}$ where $\beta_{sh} = v_{sh}/c$, and the temperature ratio of upstream ions and electrons $T_i / T_e = 1$. The widths $N_y = 11200$ and $2800$ in the Runs 7 and 8, correspond to $\approx 5$ and $1.2$ ion Larmor radii in $B_0$, respectively. The amplifications given in the last column are to the maxima in $B_\perp$ observed across several SLAMS that are closest to the shock.}
\end{deluxetable*}

We use a PIC code TRISTAN-MP (\citealt{tristan_mp}) to run simulations with low magnetizations and non-relativistic shock velocities approaching SNR conditions to study the non-linear stages of magnetic amplification by shock-accelerated particles. We run simulations in the upstream frame, using a moving reflecting piston as the left wall in an expanding simulation box. We use a gradual magnetic wall boundary condition to soften the initial cold beam reflection from the piston (for details see Sec.~\ref{sec:shkinit} in Appendix).
All simulation runs and their parameters are listed in Table~\ref{tab1}. We resolve electron inertial length (skin depth) with 10 computational cells in 1D, and with 5 cells in 2D runs. Ions and electrons are injected with the same temperatures near the right wall (in the far upstream). Since the Debye radius is not resolved, in order to reduce numerical noise we smooth the current with 16 passes of digital smoothing filter. In most of our runs we use ion-to-electron mass ratio $m_i/m_e=32$. In all runs the background field $B_0$ is aligned with the plasma flow (along $x$-axis). We first run a series of 1D high-Alfv\`en Mach number (high-$M_{\rm A}$) simulations (Runs 1--3) to check under which conditions structures similar to SLAMS observed at the Earth bow shock appear in the upstream. We find that, for Alfv\`en Mach numbers $M_{\rm A}\gtrsim 80$, the waves driven ahead of the shock get significantly amplified ($\gtrsim 10$) and turn into SLAMS. At lower Mach numbers ($20 < M_A < 40$, Runs 2-3), the amplification is smaller (factors of 3-7). We thus set $M_{\rm A} = 80$ (with $m_i/m_e=32$) as a reference case in which SLAMS can grow and evolve rapidly.

Our methodology for 2D simulations (Runs 7-8) is to first make a very wide-box run to confirm that the appearance of SLAMS is not affected by 2D geometry. Since the duration of such a run is limited by the available computational resources, we conduct a second 2D simulation with a more narrow box where we are able to push the shock evolution much further. This enables us to study particle acceleration with SLAMS in 2D at much later stages of shock evolution. Our 2D convergence runs show that there is a minimal number of particles per cell ($N_\mathrm{ppc} > 10$) needed to capture the return current and upstream waves. To ensure the growth of SLAMS and their uninterrupted evolution in 2D, we therefore use $N_\mathrm{ppc} = 32$ (or higher). Finally, we extend our 1D study to more realistic Mach numbers $M_{\rm A}=200$ (Run 5) and $M_{\rm A}=300$ (Run 6) to probe the SLAMS expected at shocks of fast and young SNRs.

\section{\label{sec:slams}The structure of SLAMS at high-$M_{\rm A}$ shocks}

\begin{figure*}[t!]\centering
\includegraphics[trim= 0px 0px 0px 0px, clip=true,width=\columnwidth]{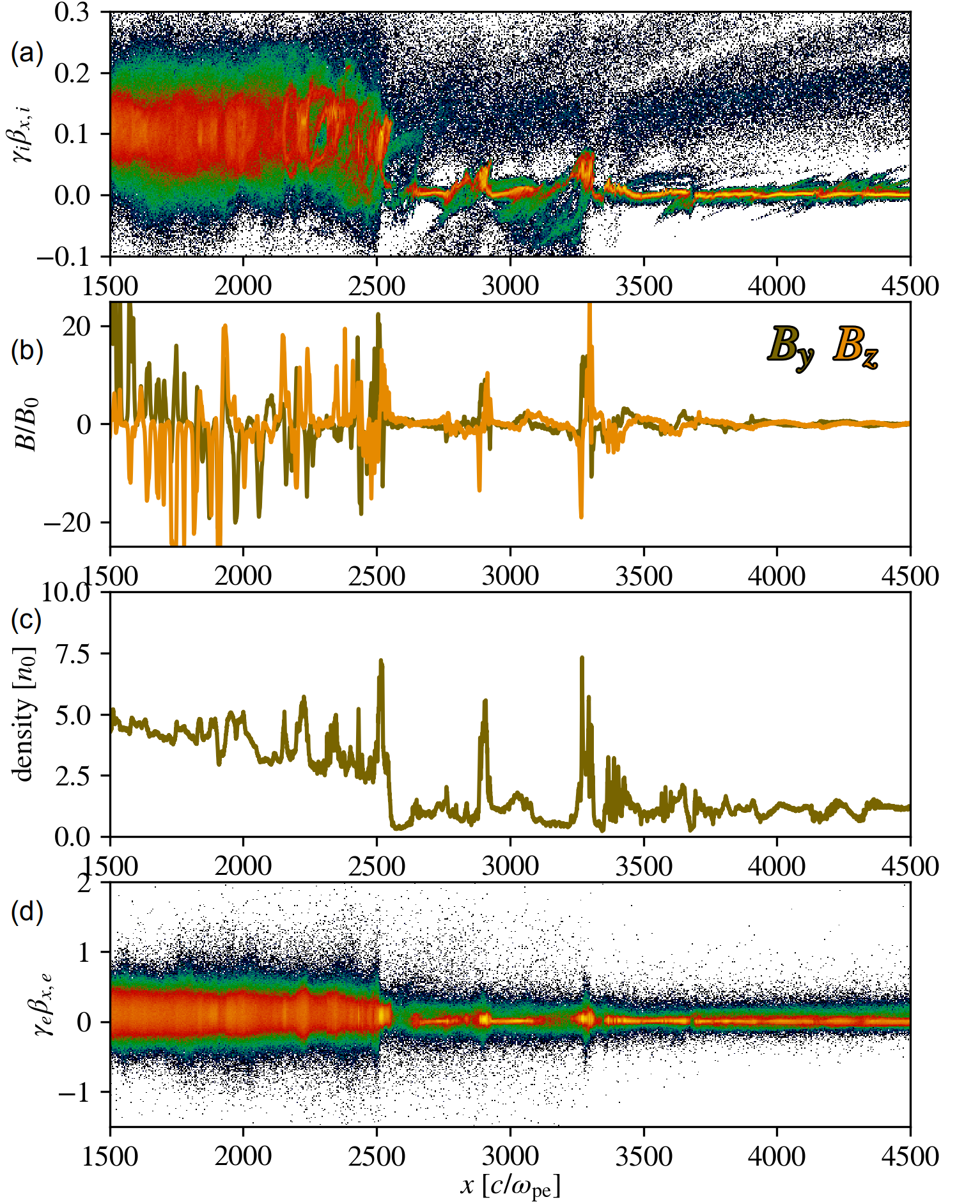}
\includegraphics[trim= 0px 0px 0px 0px, clip=true,width=\columnwidth]{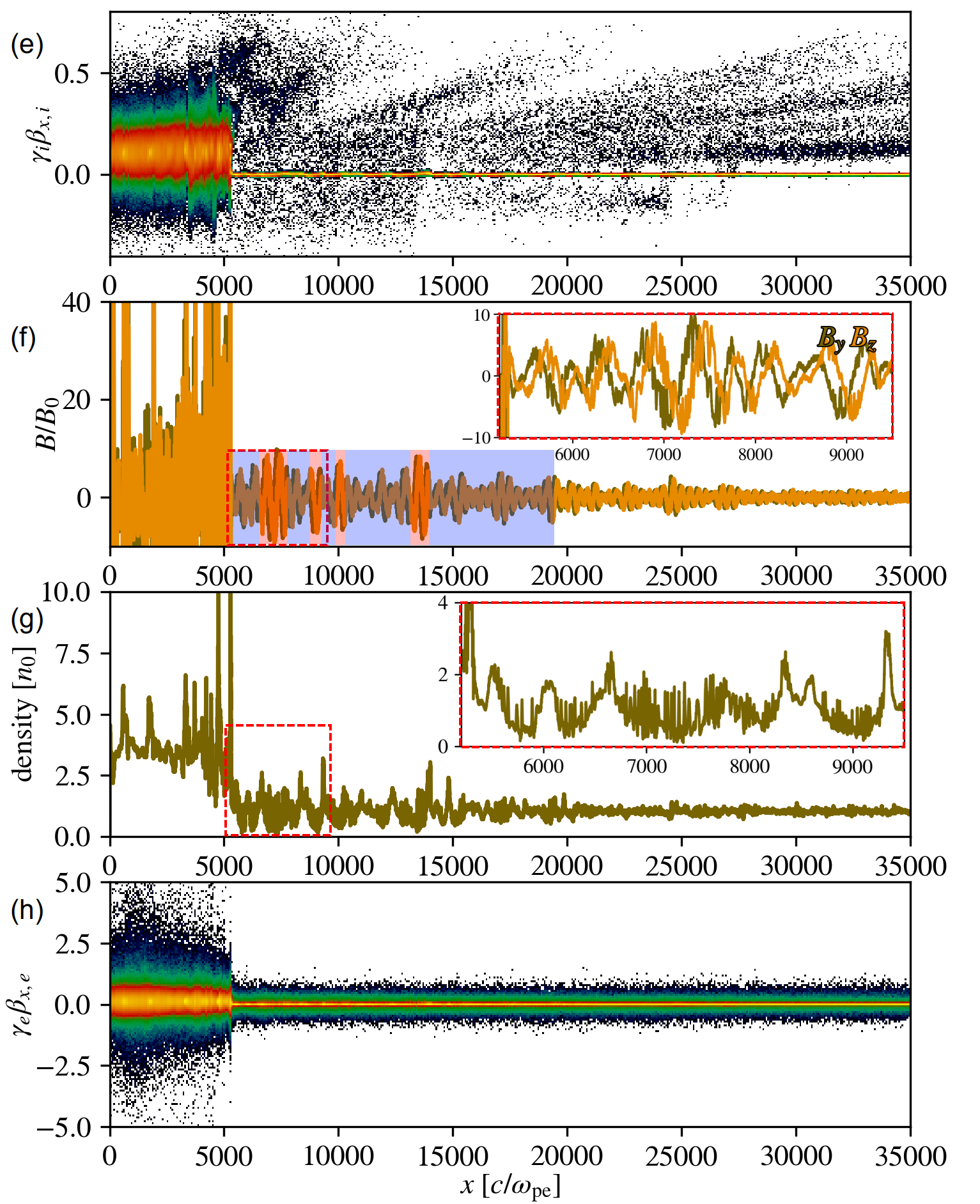}
\caption{\label{fig:1d_init} \label{fig:1d_xl}
Initial (left) and evolved (right) SLAMS in the long-term $M_{\rm A}=80$, 1D Run 1 with $m_i/m_e=32$ (1D fiducial run). From top to bottom: plots (a) and (e) show ion phase space, (b) and (f) magnetic field profiles, (c) and (g) density profile, and (d) and (h) electron phase space. The left column shows initial SLAMS at the time $t \sim 25\,\omega_{ci}^{-1}$ when they appear, and the right column shows evolved SLAMS at the end of the run $T_\mathrm{end} \sim 217\,\omega_{ci}^{-1}$. The primary (main) shock is located at $x\sim 2500\,c/\omega_{pe}$ (left) and $x\sim 5500\,c/\omega_{pe}$ (right), while in both cases quasi-periodic series of shocklets are visible in the near-upstream region. Ion diffusion can be noticed in plot (e) inside the region of $\sim 10$ developed SLAMS (i.e., $\sim 10$ AM oscillations on the $\lambda_{\rm SLAMS}$ scale) in the late stage. Subplots in (f) and (g) show a more detailed view of SLAMS and related density enhancements ahead of the shock (inside the region marked by a red rectangle). Red and blue opaque shading marks the superluminal and subluminal regions, respectively in plot (f).}
\end{figure*}

In 2D PIC simulations, a quasi-parallel high-$M_{\rm A}$ non-relativistic shock is mediated by the filamentary Weibel modes (\citealt{weibel_inst}) in the first stages of its evolution. The filaments then merge and generate the shock transition (e.g., \citealt{weibel2_inst}). Later, CR-driven modes (\citealt{CR_streaming_inst}), also known as Bell modes (\citealt{bell_inst}), grow ahead of the shock and overtake shock mediation (\citealt{weibel_to_bell}). Because of a low background magnetic field in high-$M_{\rm A}$ shocks, this process takes much longer than in the case of more magnetized (low-$M_{\rm A}$) shocks. For SLAMS to grow in the upstream, it is therefore crucial to wait for the non-resonant waves to develop ahead of the shock. In our 1D,  $M_{\rm A} = 80$ Run 1 we find that waves steepen into SLAMS after $\sim 10^5\,\omega_{pe}^{-1}$ ($\sim 30\,\omega_{ci}^{-1}$). We further recognize two types of non-linear wave structures. The early time SLAMS which are driven by the initial beam of reflected ions appear as coherent, isolated regions with enhanced magnetic field. This closely matches the original definition of SLAMS as given in \cite{SLAMS_init}, \cite{SLAMS_obs_init}, and \cite{Lucek2002}. Later on, as the precursor of diffusive CRs forms, a smooth oscillating pattern modulates the upstream waves on the scales that are much longer than the periodicity of early SLAMS. Similar patterns are observed in the Earth's foreshock as amplitude-modulated (AM) ultra-low-frequency (ULF) waves~\cite{Eastwood2005}. However, the large-period AM patterns in our runs appear with $\sim 10$ times larger spatial scale than ULF waves (which scale is comparable to the scale of the carrier wave in our runs). Such disagreement is likely to appear because we simulate shocks in a high-Mach number regime where the Larmor radius of returning ions is proportionally larger than that of ions reflected at the low-Mach number Earth's bow shock. We refer to the large-period oscillations that we observe in our runs as developed SLAMS instead of ULF waves or whistler precursor.

In Fig.~\ref{fig:1d_init} we show the early (left column) and late (right column) stages of the shock evolution from Run 1. The initial ion beam with temperature $k_\mathrm{B} T / m_e c^2 \sim 0.13$ (where $k_\mathrm{B}$ is Boltzmann constant) and drift velocity $\sim 1.5\,v_{sh} = 0.2\,c$ drives non-resonant Bell waves on a scale $\lambda_{\rm CR} \sim 150-200\,c/\omega_{pe}$ that is initially smaller than the Larmor radius of reflected ions in the region $x\sim 3400-4500\,c/\omega_{pe}$ in Fig.~\ref{fig:1d_init}b. Early or initial SLAMS appear as two short, large-amplitude pulses of transverse magnetic field, as can be seen in the region $x\sim 2600-3400\,c/\omega_{pe}$. The pulses are induced by the collective motion of returning ions, which appears as ion loops in the same region in phase space in Fig.~\ref{fig:1d_init}a (for details see Sec.~\ref{sec:shkinit} in Appendix), followed by the heating of upstream electrons at the pulse maxima (see electron phase space in Fig.~\ref{fig:1d_init}d).
\begin{figure*}[t!]
\centering
\includegraphics[trim=15px 0px 10px 0px, clip=true,width=0.53\textwidth]{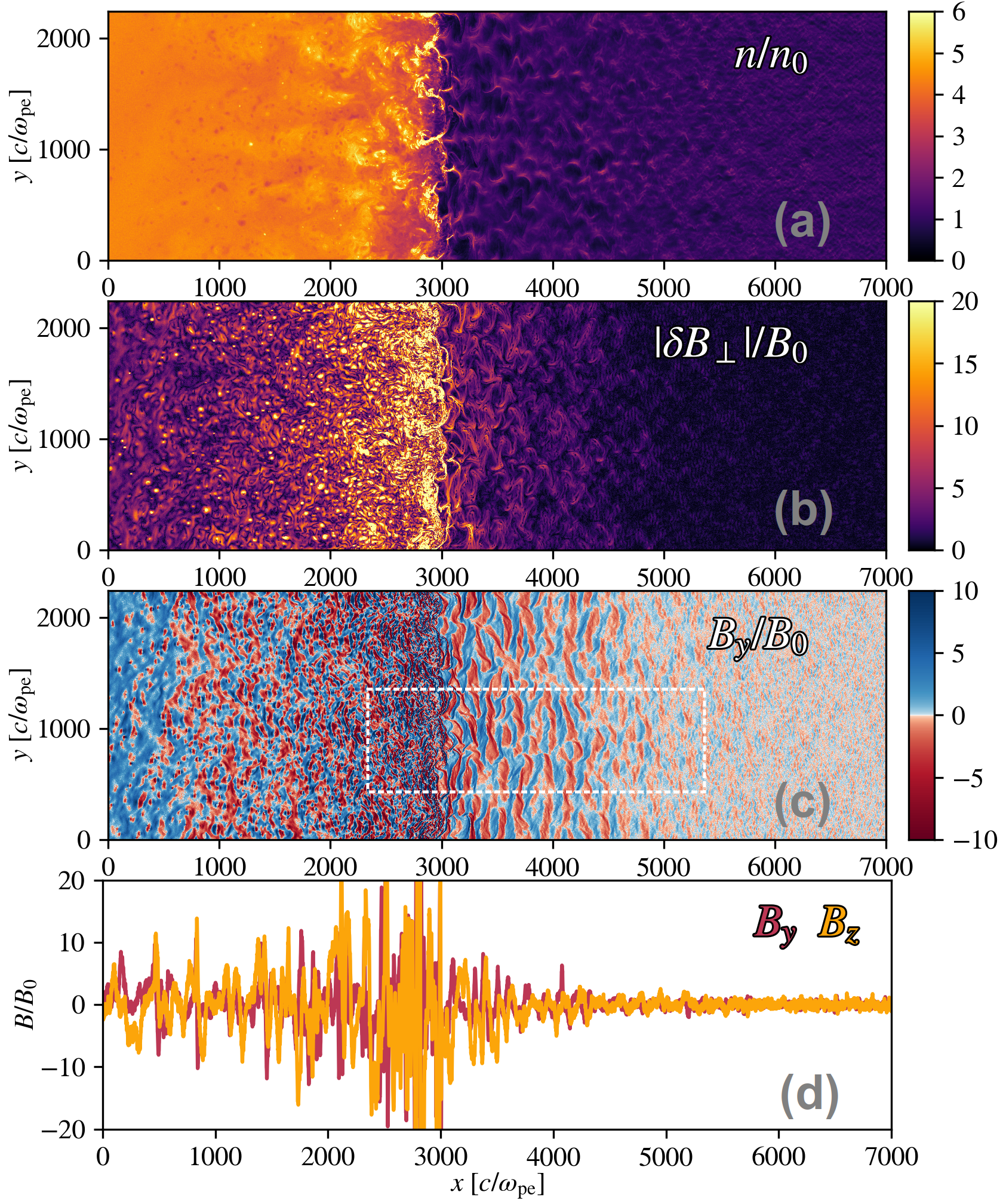}
\includegraphics[trim=60px 0px 180px 0px, clip=true,width=0.461\textwidth]{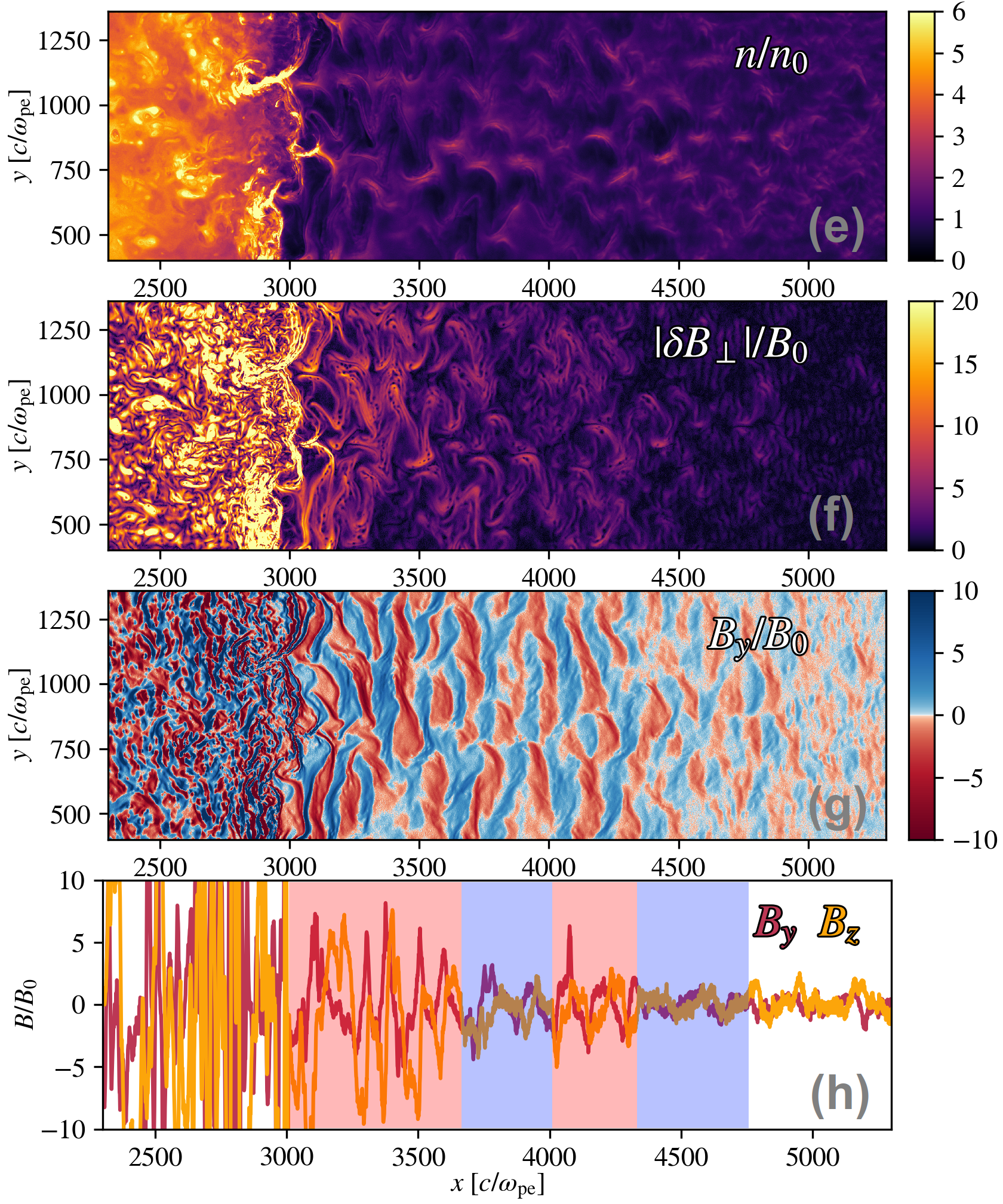}
\caption{\label{fig:2d_w11200_frontera_init} 
Early SLAMS in the large box $M_{\rm A}=80$ Run 7 ($m_i/m_e=32$, $N_\mathrm{ppc}=32$) at $t \sim 60\,\omega_{ci}^{-1}$. From top to bottom: plots (a) and (e) show the density map, (b) and (f) total perpendicular magnetic field, (c) and (g) in-plane magnetic field, and (d) and (h) transverse field profiles (averaged over a narrow $y$-region) with red and blue opaque shading of the superluminal and subluminal regions, respectively. The diagrams in the left column cover the entire transverse size of the simulation domain (while longitudinally it is not shown entirely). The plots in the right column show the enlarged central region that is marked by a white rectangle in plot (c), which encompasses a small portion of the downstream, the shock at $x\sim 3000\,c/\omega_{pe}$, and the precursor.}
\end{figure*}
Such initial SLAMS are transients since they appear as very strong only during the initial phases of shock evolution. At their saturation, the initial SLAMS move away from the shock at a constant velocity $v_{\mathrm{SLAMS}} \sim v_{\mathrm{sh}}/4 -v_{\mathrm{sh}}/2$ in the upstream frame. SLAMS thus drive strong transient shocks that propagate along $x$-axis at Mach numbers $\sim 25-40 < M_\mathrm{A}$ in the upstream, with density overshoots exceeding 4 (e.g., $x\sim 2700-3500\,c/\omega_{pe}$  in Fig.~\ref{fig:1d_init}c).

\begin{figure*}[t!]
\centering
\includegraphics[trim=0px 0px 0px 0px, clip=true,width=\textwidth]{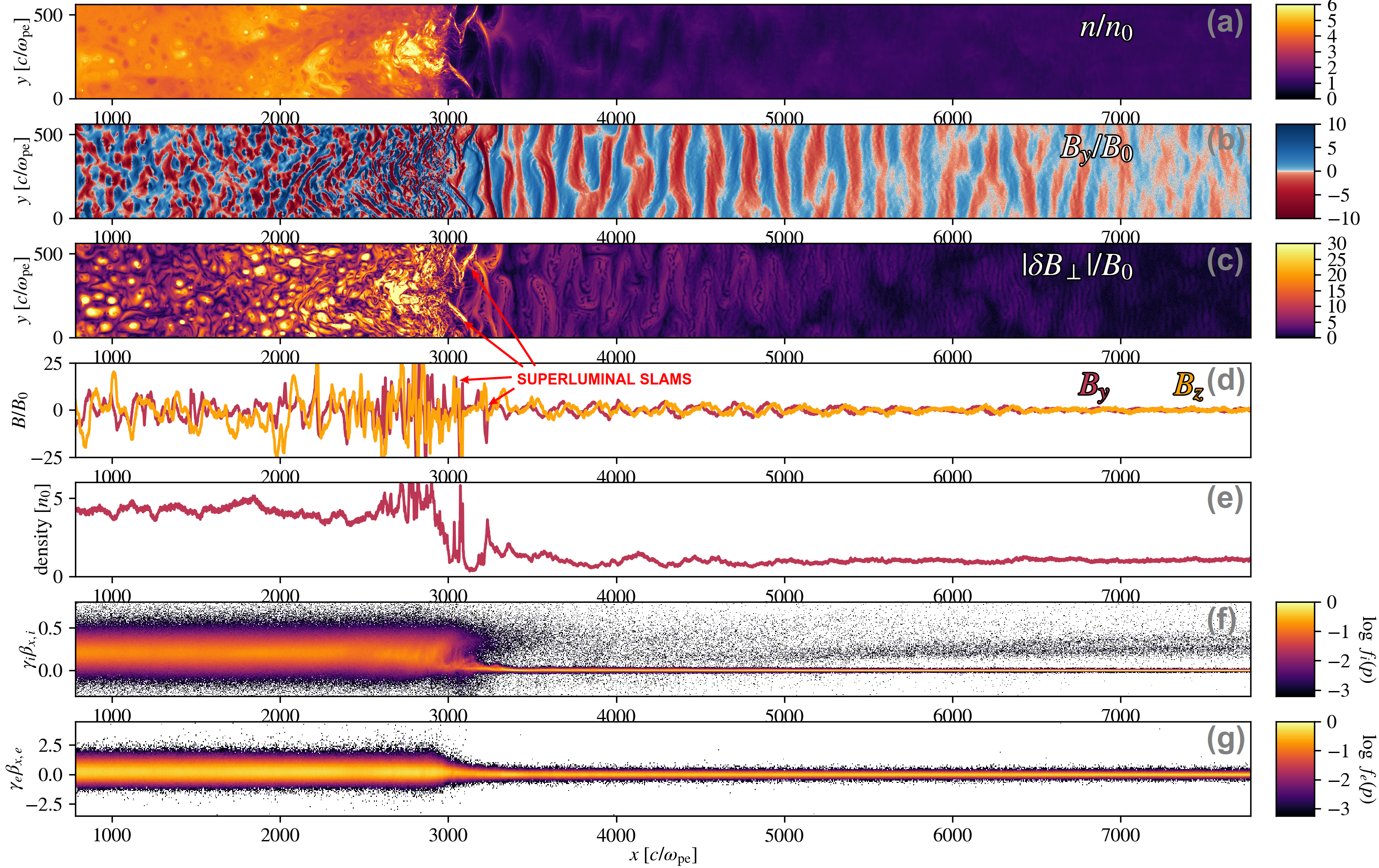}
\caption{\label{fig:2d_w2800_perl} 
Evolved, superluminal SLAMS supported by the self-consistently accelerated CRs in the 2D Run 8 at the end time $T_{\rm end} \sim 204\,\omega_{ci}^{-1}$. From top to bottom: (a) density map, (b) in-plane magnetic field, (c) total perpendicular magnetic field, (d) transverse field profiles (averaged over a narrow $y$-region), (e) density profile, (f) ion phase space, and (g) electron phase space. The primary shock is located at $x\sim 2800\,c/\omega_{pe}$, while
the new shock fronts are visible as tiny filaments in the region $x\sim 3000-3200\,c/\omega_{pe}$ in plots (a--c) associated also with the density spikes in plot (e). They form due to a super-Alfv\`enic push which SLAMS exert on the upstream plasma. The shocklets grow from the field and density enhancements farther ahead as in plots (d) and (e), respectively.}
\end{figure*}

SLAMS in the late stages of evolution (at $\sim 5-6 \times 10^5\,\omega_{pe}^{-1}$ or $\sim 150-180\,\omega_{ci}^{-1}$), shown in Fig.~\ref{fig:1d_xl}e-h, are driven by a diffuse beam of accelerated CRs (with temperature equivalent of $k_\mathrm{B} T / m_e c^2 \sim 1.12$ and drift velocity $\sim v_{sh} = 0.133\,c$), rather than by a coherent beam of reflected particles. CR ions are able to diffuse and fill the phase space nearly uniformly in the region $x\sim 5000-15000\,c/\omega_{pe}$ in Fig.~\ref{fig:1d_xl}e. Contrary to the coherent gyrations of returning ions which make the pulse-shaped initial SLAMS, CR ions induce a periodic large-scale amplitude modulation (AM) on self-driven non-resonant waves (Fig.~\ref{fig:1d_xl}f). We refer to the wave packets such as the one highlighted by the red rectangle in the region $x\sim 5000-8500\,c/\omega_{pe}$ in Fig.\ref{fig:1d_xl}f, as the developed or evolved SLAMS. We detect the appearance of several spatial scales of AM -- $\lambda_{\rm SLAMS} \sim 500,\,1000\,\&\,2000\,c/\omega_{pe}$ or $\sim 1,\,2\,\&\,4\,r_\mathrm{L,sh}$, as multiples of the Larmor radius $r_\mathrm{L,sh} = M_\mathrm{A} \sqrt{m_i/m_e}$ of ions drifting with $v_\mathrm{sh}$ in $B_0$. All three scales increase over time as the average kinetic energy of CR ions grows. Since the density of the driving beam gradually decreases with time, $\lambda_{\rm CR}$ increases as well. Due to the low density of accelerated CR ions ($n_{\rm CR}/n_0 \sim 0.01-0.05$), the transient shocks driven by evolved SLAMS become much weaker farther in the upstream (propagating at Mach numbers $\sim M_{\rm A}/10$). The upstream density enhancements in Fig.~\ref{fig:1d_xl}g, which are associated with SLAMS, show compression $< 4$. We observe that the CR precursor with developed SLAMS spreads into the upstream at a roughly constant rate (which is higher than the advection rate).

We find that SLAMS with similar properties grow and evolve in the case with a more realistic mass ratio $m_i/m_e=100$ in 1D Run 4. We confirm that both the initial and developed SLAMS in the case with $m_i/m_e=100$ show similar growth rates, amplification, and scales $\lambda_{\rm CR}$ and $\lambda_{\rm SLAMS}$ as the SLAMS in runs with $m_i/m_e=32$ (see Sec.~\ref{sec:mime100} in Appendix).

Guided by the results of 1D runs, we use the same parameters for our 2D runs, choosing $N_{\rm ppc}\geq 32$ based on the convergence studies described in Sec.~\ref{sec:setup}. To capture 2D dynamics of SLAMS and their associated electron acceleration, we perform a simulation with large transverse size in Run 7. In Fig.~\ref{fig:2d_w11200_frontera_init} we show the density map in panels (a,e), total perpendicular field $B_\perp$ in (b,f), in-plane component $B_y$ in (c,g), and average $B_{y,z}$ profiles in (d,h). The left column shows the full transverse size of the simulation, while the right column shows a zoomed-in region near the shock (shown as white rectangle in \ref{fig:2d_w11200_frontera_init}c). In the field maps in Fig.~\ref{fig:2d_w11200_frontera_init}c,g we notice the well-organized transverse field $B_y$ of initial SLAMS with $\lambda_{\rm SLAMS}\sim\lambda_{\rm CR}$ that drives density cavities of the same size in the upstream (Fig.~\ref{fig:2d_w11200_frontera_init}a,e). We find that the cavities are correlated with the amplified $B_\perp$-filaments (compare Fig.~\ref{fig:2d_w11200_frontera_init}e and f), meaning that plasma is evacuated from these regions by magnetic pressure gradients. As SLAMS-driven cavities and amplified filaments are advected to the shock, they corrugate the shock surface. Although the upstream density looks quite turbulent, the $B_y$ field seen in Fig.~\ref{fig:2d_w11200_frontera_init}c,g only changes its phase along $y$-axis. In Fig.~\ref{fig:2d_w11200_frontera_init}a,b,e, and f we observe the merging of density cavities and the formation of serpentine patterns that are connected to the gyration of returning ions. The patterns look very similar to those in the case of $M=100$ hybrid run in \citet{DamAnatoly2014}. Compared to the 1D case, early SLAMS in Run 7 grow on similar time and spatial scales, with the amplification reaching about the same level ($B_\perp/B_0\sim 10$ close to the shock).

Besides the Bell-scale SLAMS (with $\lambda_{\rm SLAMS}\sim \lambda_{\rm CR}$), we observe the appearance of large-scale AM oscillation in 2D (with $\lambda_{\rm SLAMS}\sim 5\,\lambda_{\rm CR} \sim 700-800\,c/\omega_{pe} \sim 2\,r_\mathrm{L,sh}$) visible in $B_{y,z}$ profiles at $x\sim 3000-5000\,c/\omega_{pe}$ in Fig.~\ref{fig:2d_w11200_frontera_init}d,h. Similarly to the short-scale SLAMS with $\lambda_{\rm SLAMS}\sim \lambda_{\rm CR}$, the large-scale SLAMS are also very coherent. This is very important because the periodicity of the large-scale SLAMS defines the maximum energy that accelerating electrons can reach as we show in the next section.

Reaching late stages in 2D with the available computational resources is only possible with the use of a smaller transverse size of the simulation box.
To study the developed SLAMS in 2D, we make Run 8 where we use most of the parameters from Run 7, but decrease the transverse size to capture only a few cavities. In Fig.~\ref{fig:2d_w2800_perl} we present the density, $B_y$ and $B_\perp$ field maps in plots (a--c), the averaged $B_{y,z}$ and density profiles in plots (d,e), and ion and electron phase spaces in plots (f,g) at the end of Run 8. Similar to the previous cases, the developed SLAMS appear on two scales $\lambda_{\rm SLAMS} \sim 500\,\&\,1000\,c/\omega_{pe} \sim 1\,\&\,2\,r_\mathrm{L,sh}$. The scales that are visible as AM structures in early time in Fig.~\ref{fig:2d_w11200_frontera_init}d,h are later distributed across the larger portion of the upstream in Fig.~\ref{fig:2d_w2800_perl}d. We observe these SLAMS become significantly amplified as they get close to the shock. SLAMS then drive shocklets that often appear as thin membranes in density and $B_\perp$ with an irregular or curved shape in the region $x\sim 3000-3200\,c/\omega_{pe}$ in plots Fig.~\ref{fig:2d_w2800_perl}a,c, associated with the spikes in integrated density profile in plot (e).
During the run we observe that the SLAMS-driven cavities and associated serpentine patterns in $B_\perp$ cascade to larger sizes. By the end of the run ($T_{\rm end}=204\,\omega_{ci}^{-1}$) these structures reach the transverse size of the simulation box as shown in Fig.~\ref{fig:2d_w2800_perl}a,c.

The near upstream region in phase space in Fig.~\ref{fig:2d_w2800_perl}f is populated by the diffusing CR ions. The colder ion beams observed farther in the upstream ($x > 5000\,c/\omega_{pe}$) represent the relic of the previous strong shock reformation induced by SLAMS. We find that such a time variability is common to SLAMS even during the late stages of their evolution. It can manifest through sudden and strong shock reformations, or through variability (appearance/disappearance) of the AM envelope of SLAMS on long time scales. Despite this time variability, we observe that SLAMS continuously build up upstream of the shock and maintain roughly the same amplification throughout their evolution. The dynamics of SLAMS is highly non-linear, and at higher Mach numbers we expect SLAMS to show even more oscillatory behavior. Nevertheless, our study implies that SLAMS are indeed persistent and not a transient phenomenon in high-$M_{\rm A}$ shocks.

\section{\label{sec:elacc}Electron acceleration with SLAMS}

Even though in the late stages in 2D only the first few SLAMS in the precursor reach the amplification as high as in 1D and can become superluminal, we show in this section that this is sufficient to make SLAMS extremely fast electron accelerators. Despite the strong electron advection induced by SLAMS, we observe very fast formation of a non-thermal tail in the electron spectrum behind the shock. On the other hand, ions develop only a short non-thermal tail. Since SLAMS are driven on the ion scales, we observe a clear diffusion of ions in the upstream. However, CR ions accelerating via DSA do not reach high energies by the end of our runs, which sets our main focus to electron acceleration. In this section we present the electron spectra from 1D and 2D runs and show the trajectories of accelerated electrons. We introduce a new mechanism for electron acceleration at high-$M_{\rm A}$ shocks and provide a possible theoretical interpretation for the electron spectra. Finally, we apply our model to explain the spectra observed in young SNRs.

\begin{figure}[t!]\centering
\includegraphics[trim=0px 0px 0px 0px, clip=true,width=\columnwidth]{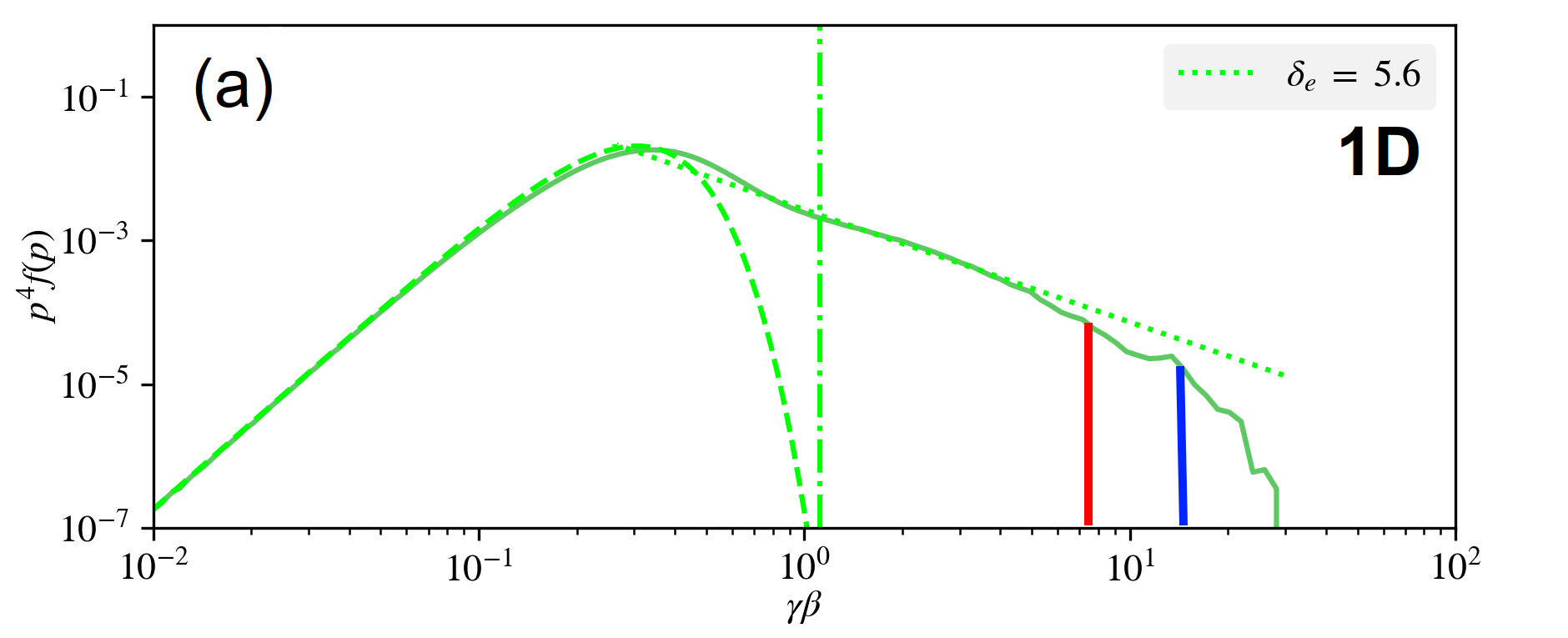}
\includegraphics[trim=0px 0px 0px 0px, clip=true,width=\columnwidth]{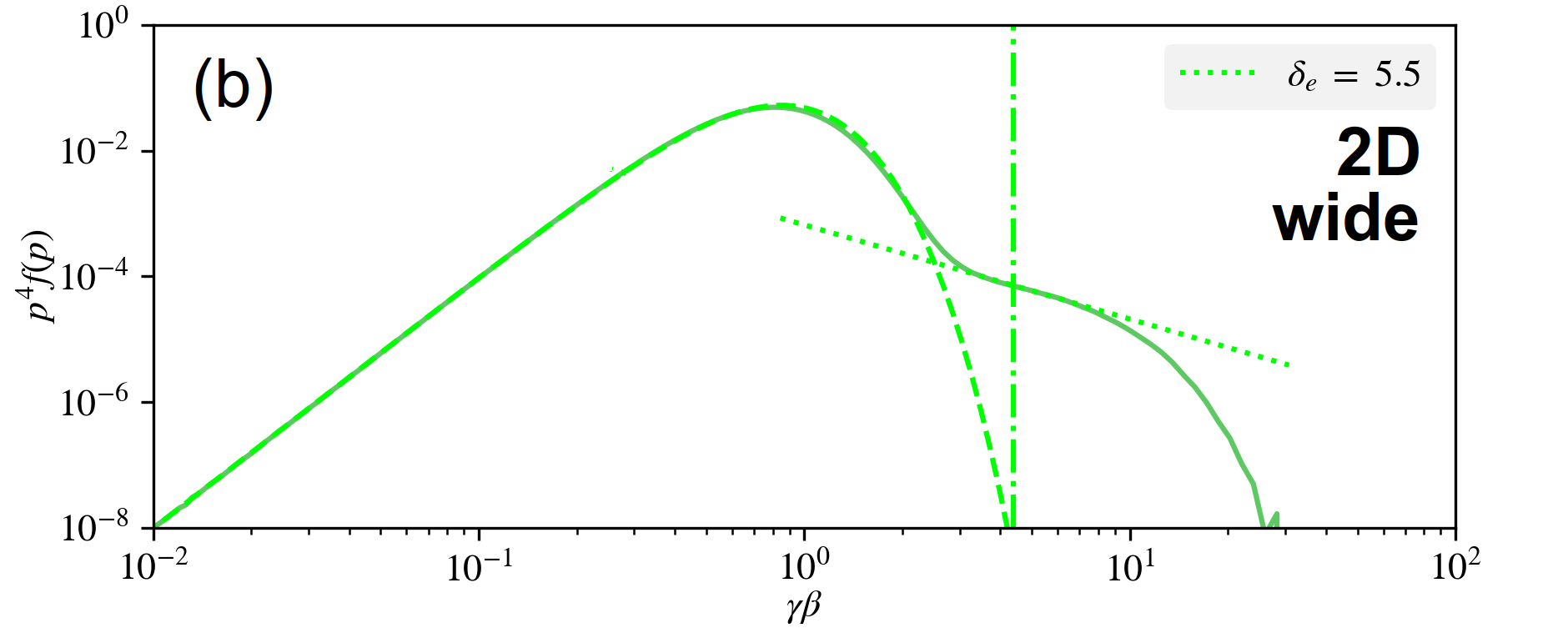}
\includegraphics[trim=0px 0px 0px 0px, clip=true,width=\columnwidth]{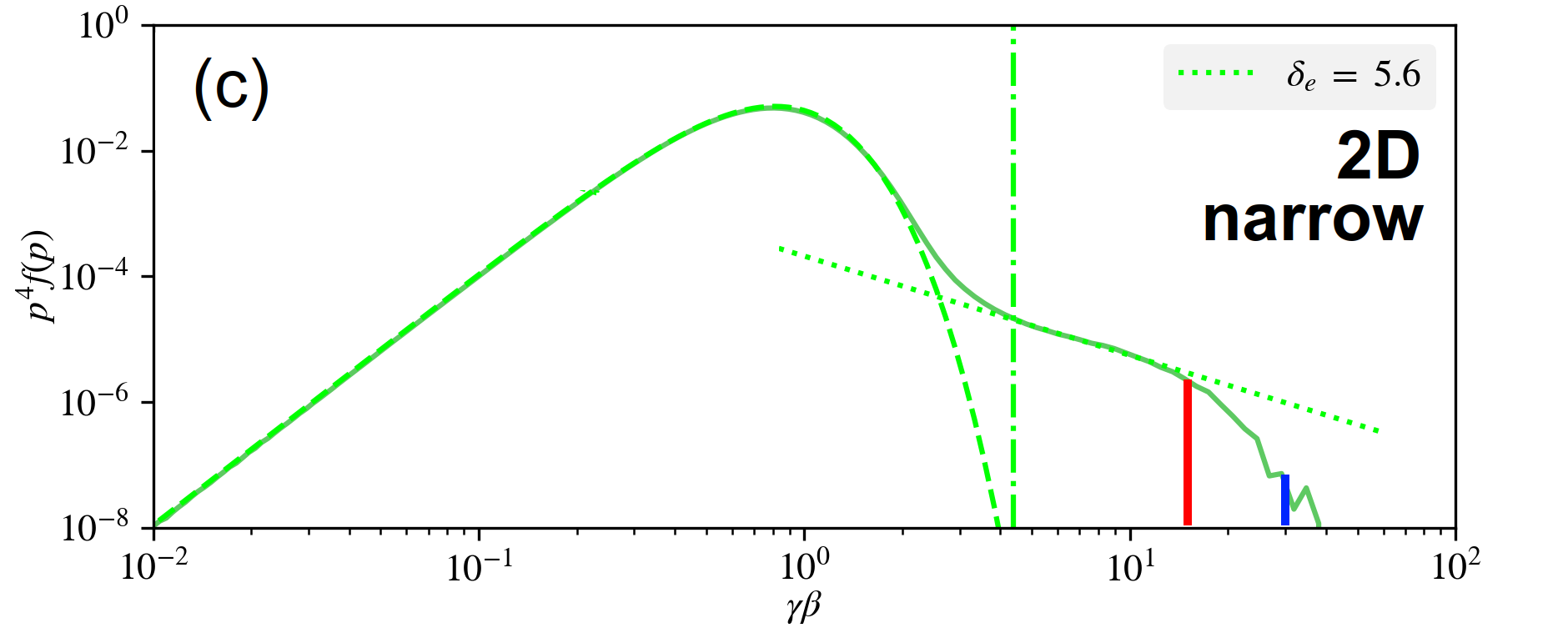}
\caption{\label{fig:spectra} 
The downstream electron spectra in the case of $M_{\rm A}=80$ in the 1D fiducial Run 1 (top), wide 2D Run 7 (middle), and narrow 2D Run 8 (bottom) at the final times $217~\omega_{ci}^{-1}$, $60~\omega_{ci}^{-1}$, and $204~\omega_{ci}^{-1}$, respectively. The red and blue vertical lines mark the momenta where electrons are in resonance with the two large SLAMS' scales that shape the wave envelope --- $\lambda_{\rm SLAMS} \sim 500~\&~1000~c/\omega_{pe}$, respectively. The vertical dot-dash lines mark the momenta where QSA tail begins to dominate over Maxwellian.}
\end{figure}

\subsection{\label{sec:spectra}Electron spectra and acceleration mechanism}

\begin{figure*}[t!]\centering
\includegraphics[trim=2px 2px 2px 2px, clip=true,width=\columnwidth]{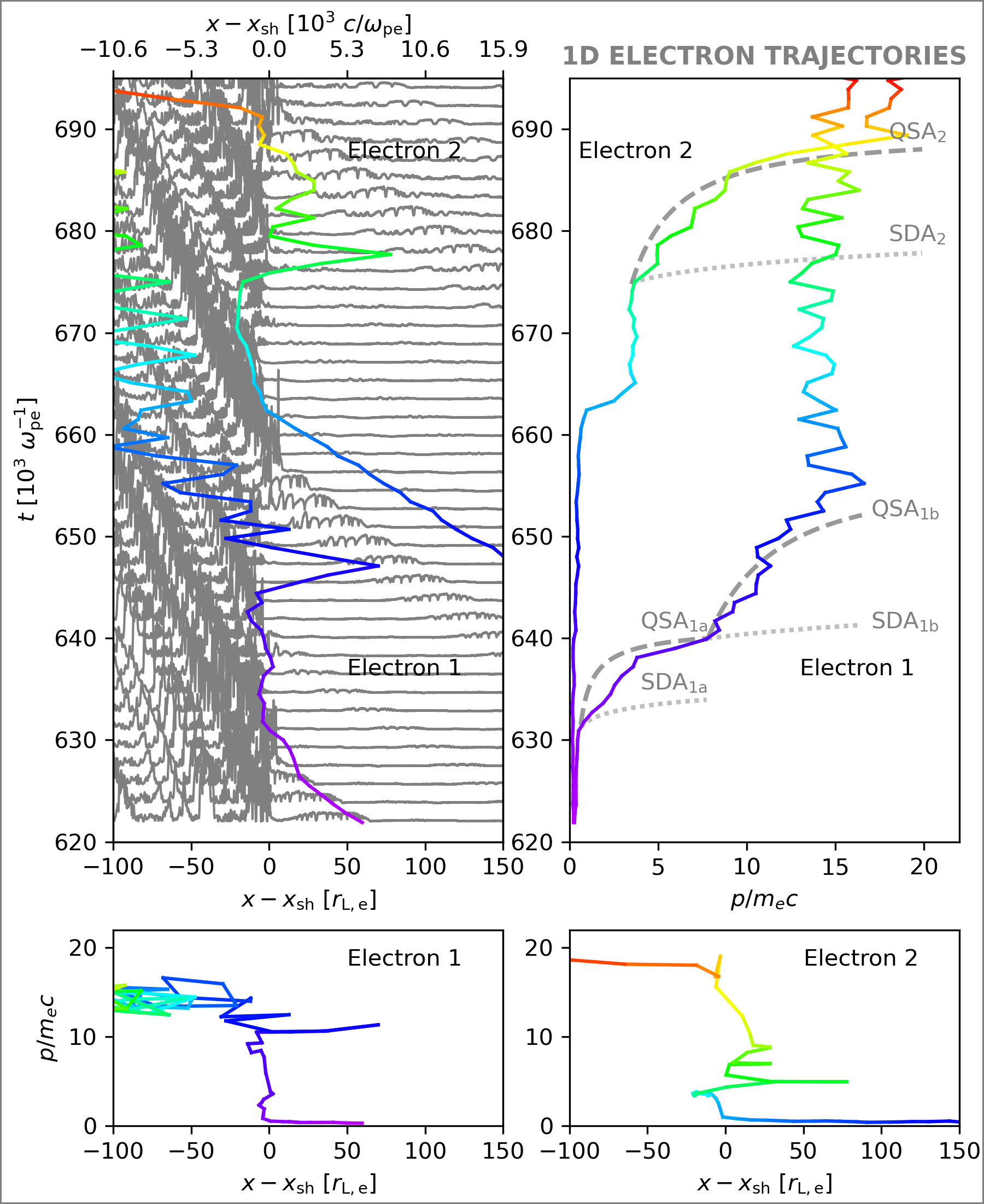}
\includegraphics[trim=2px 2px 2px 2px, clip=true,width=\columnwidth]{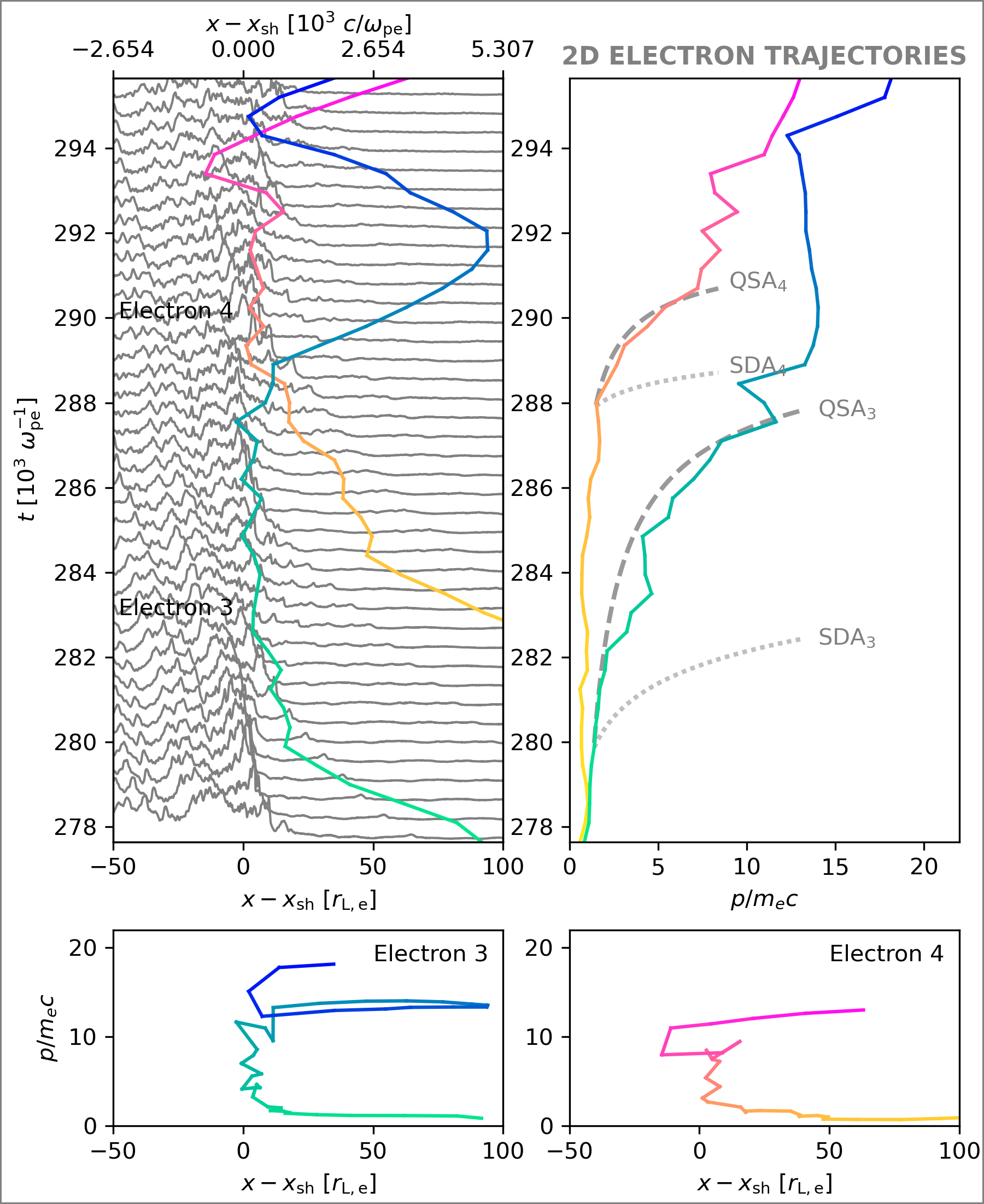}
\caption{\label{fig:elec_plots} 
Trajectories of representative high-energy electrons from the 1D fiducial Run 1 (left panels) and 2D Run 8 (right panels). In each case the panels show trajectories of two electrons in $x-t$ space (top left) and in total momentum space (bottom panels). Growth of the particle momenta over time is compared with theoretical predictions in the top right panels for both runs. The predictions, derived in Sec.~\ref{sec:accrate} of the Appendix, are shown with dashed lines for Fermi I or QSA (Eq.~\ref{equ:tqsa}) and dotted lines for SDA (Eq.~\ref{equ:tsda}) mechanisms. Different colors denote the time along particle trajectory. The spatial coordinates are given relative to the shock location (at $x=0$) in the units of electron Larmor radius $r_{\rm L,e}$ for electrons moving with $v=v_\mathrm{sh}$ in $B_0$ (bottom axes) and electron skin depth $c/\omega_{pe}$ (top axes). The time is in units of inverse plasma frequency $\omega_{pe}^{-1}$. The $|B_\perp|$-profiles at specific times are plotted in gray in the background.}
\end{figure*}

In Fig.~\ref{fig:spectra}, we show the downstream electron spectra in our 1D and 2D PIC runs where SLAMS reach the evolved stage. In all cases we find that the electron non-thermal tail develops very fast once SLAMS appear in the upstream. The non-thermal tail remains steepened by a power of 1 to 2 in momentum relative to the DSA prediction of $\delta_e = 4$ in electron momentum distribution $f(p) \sim p^{-\delta_e}$. During the early stages of strong initial SLAMS we observe $\delta_e \sim 6$. As the amplification of SLAMS settles at later stages, the momentum index relaxes to slightly lower values which are close to $\delta_e \sim 5.0-5.5$. At the end of Runs 1 (1D) and 8 (2D) electrons reach the resonant energies where their Larmor radii in the amplified field are $r_{{\rm L},e} \sim \lambda_{\rm SLAMS} /2\pi$ implying the resonant momentum
\begin{eqnarray}
    \frac{p_{{\rm res},e}}{m_e c} &&= \frac{r_{{\rm L},e} \cdot \omega_{ce}}{c} \frac{B}{B_0} = \frac{\lambda_{\rm SLAMS}}{2\pi} \cdot \frac{\omega_{pe}}{c}\sqrt{\sigma_e} \, \frac{B}{B_0}\nonumber \\
    &&= \frac{\lambda_{\rm SLAMS}}{2\pi\,r_{{\rm Lsh},i}}\cdot  \beta_{sh} \, \frac{m_i}{m_e} \, \frac{B}{B_0}, \label{equ:res}
\end{eqnarray}
where $\omega_{ce}$ is the cyclotron frequency of a non-relativistic electron in $B_0$, $\sigma_e$ is the electron magnetization in the base field $B_0$, $r_{\mathrm{L},e}$ is calculated for a total field $B = \sqrt{B_\perp^2 + B_0^2}$, $r_{{\rm Lsh},i} = M_\mathrm{A} \sqrt{m_i/m_e} \cdot c/\omega_{pe}$ is the Larmor radius of ions moving with $v_\mathrm{sh}$ in the base field $B_0$, and $\beta_{sh} = v_{sh}/c$.
We use Eq.~\ref{equ:res} to calculate the resonant momenta for the two most dominant SLAMS scales in the runs ($\lambda_{\rm SLAMS} \sim 500\,\&\,1000\,c/\omega_{pe}$) and mark these momenta with the red and blue lines in Fig.~\ref{fig:spectra}a,c, respectively. The most energetic electrons are able to reach the highest resonant momentum (the blue line) except in the wide-box Run 7 in Fig.~\ref{fig:spectra}b where SLAMS are not yet as developed as in the long-term runs. At the end of our longest simulation (Run 1 in Fig.~\ref{fig:spectra}a) we observe the appearance of an even larger scale $\lambda_{\rm SLAMS} \sim 2000\,c/\omega_{pe}$ (visible in Fig.~\ref{fig:1d_xl}f) which implies $p_{{\rm res},e}/m_e c \sim 30$. We observe that magnetized electrons reach $p_{{\rm res},e}$ during the advection period of a single resonant upstream structure, which is $\tau_\mathrm{adv} \sim 0.01\, T_\mathrm{end}$. However, for an unmagnetized electron with $p \sim 10\, p_{{\rm res},e}$ it would take at least 10 times longer ($\gtrsim 0.1\,T_\mathrm{end}$) to complete a single DSA cycle. This agrees with our observation that only a small fraction of unmagnetized electrons reach $p > p_{{\rm res},e}$.

To understand the fast electron acceleration we pick four electrons from the high-energy part of the spectra in 1D and 2D runs and analyze their trajectories (shown in Fig.~\ref{fig:elec_plots}). We find that electrons gain most of their energy right in front of the shock ($0 < x < \lambda_{\rm SLAMS}$).
Electrons bounce between the shock and the nearest superluminal SLAMS' maximum, e.g., the one positioned at $x-x_{\mathrm{sh}} \sim 150\ r_{\mathrm{L},e}$ for $t \sim 640 \times 10^3\,\omega_{pe}^{-1}$ (see $x-t$ plot in 1D case in Fig.~\ref{fig:elec_plots}). As SLAMS are advected towards the shock, the reflections which are indicative of Fermi I cycles become more frequent and result in shrinking bounce patterns in electron trajectories. The 1D trajectories of both Electrons 1 and 2 show segments with such patterns in $x-t$ space at periods $t \sim [640 - 655] \times 10^3\,\omega_{pe}^{-1}$ and $[675 - 690] \times \,10^3\,\omega_{pe}^{-1}$, and in momentum space at $p/m_e c \sim 8-15$ and $4-18$, respectively. At the end time of 2D Run 8 (right plots) SLAMS are not yet as developed as in 1D, and a shrinking pattern can clearly be identified only in the short segment of trajectory of Electron 4 at $t \sim [288 - 291] \times 10^3\,\omega_{pe}^{-1}$ in $x-t$ plot and $p/m_e c \sim 2-8$ in momentum space. However, we find a much better indication of this peculiar Fermi I acceleration in the corresponding trajectories in $p-t$ space (top right plots in both 1D and 2D cases). We observe that the momentum grows over time in a fast and non-linear manner, untypical of either SDA or DSA. The momentum growth agrees well with the theoretical prediction\footnote{The prediction for the growth of momentum with time in the case of Fermi I mechanism is represented by Eq.~\ref{equ:tqsa} which is later derived in Sec.~\ref{sec:accrate} in Appendix.} for the Fermi I mechanism (dashed curves labeled as $\mathrm{QSA}_\mathrm{[1b,2,3,4]}$) that we describe in Sec.~\ref{sec:qsa} and derive in Sec.~\ref{sec:qsamod}. The momentum growth thus significantly deviates from the SDA prediction\footnote{The SDA prediction is represented by Eq.~\ref{equ:tsda} which is derived in the case of (quasi-)parallel shocks in Sec.~\ref{sec:accrate} in Appendix.} (dotted curves labeled as $\mathrm{SDA}_\mathrm{[1b,2,3,4]}$). The patterns in $x-t$ and $p-t$ trajectories that we find in our runs are quite similar to the patterns observed in the trajectories of particles that accelerate by reflecting from oppositely propagating MHD solitons \citep{Kuramitsu_Hada_2000}. In both cases the acceleration mechanism is Fermi I, with particles performing mirror reflections from nonlinear magnetic structures and experiencing a nearly free flight in between the reflections. In our runs, we find that electrons collectively enter such cycles which get synchronized by the advection period of the largest SLAMS scale. We explain the details of this acceleration mechanism and give an analytical model in the following sections.

Vertical structure which appears at $p/m_e c \sim 1-4$ in the momentum space of Electron 2 reveals that  Fermi I acceleration can be preceded by SDA cycles at the shock transition in 1D. The corresponding SDA portion in $p-t$ trajectory appears as a short segment located at $t \sim 663 \times 10^3\,\omega_{pe}^{-1}$ with the slope similar to the SDA prediction (dotted curve $\mathrm{SDA}_\mathrm{1a}$\footnote{Since both Electron 1 and Electron 2 start to accelerate at the same momentum, the same SDA curve is used to compare their initial momentum growth.}). The portion of the Electron 1 trajectory $p/m_e c \sim 4-8$ at $t \sim 640 \times 10^3\,\omega_{pe}^{-1}$ also agrees with the same SDA prediction. However, in the preceding part $p/m_e c \sim 1-4$, the curve shows a slower growth with a break at $p/m_e c \sim 4$. Since the slope of Fermi I curve (dashed line labeled as $\mathrm{QSA}_\mathrm{1a}$) is similar to that of $\mathrm{SDA}_\mathrm{1a}$ at $p/m_e c \sim 4-8$, it remains unclear whether it is indeed SDA, or the whole portion $p/m_e c \sim 1-8$ is a combination of SDA and Fermi I mechanisms. In the 2D case, such SDA portions in electron trajectories do not appear as frequently as in 1D, which may be caused by transverse diffusion of electrons in 2D. Instead, we find that some electrons pre-accelerate while being advected across the shock precursor formed by the strong initial ion beams during early stages in 2D Run 8. We observe that electrons bounce between the converging wave fronts due to the existence of a velocity gradient along the precursor itself, as predicted by \cite{precursor_acc}. 
However, pre-acceleration by both SDA and precursor mechanisms does not result in energies much higher than initial electron energies, so they do not contribute to high energy electron tail in our runs.
The periodic and collective Fermi I acceleration (which corresponds to the segments of electron trajectories fitted by dashed curves in $p-t$ plots in Fig.~\ref{fig:elec_plots}) thus dominates later, at higher electron energies in all our 1D and 2D runs. It is interesting that even though the density cavities and field patterns associated with SLAMS have 2D structure (as in Fig.~\ref{fig:2d_w11200_frontera_init}), the trajectories and spectra of electrons remain quite similar to the 1D case.

\subsection{\label{sec:qsa}Quasi-periodic shock acceleration}

In Sec.~\ref{sec:slams} we showed that in the case $M_{\rm A} \gtrsim 80$, SLAMS amplify $B_\perp$ by a factor of $\gtrsim 10$ (which we find to increase with the shock Mach number). Also, we find that scales $\lambda_{\rm SLAMS}$ and $\lambda_{\rm CR}$ are much larger than the Larmor radius of the upstream thermal electrons. Such an amplified large scale magnetic field thus significantly alters electron injection, diffusion, and acceleration. In order to propagate upstream by gliding along highly inclined magnetic field lines, adiabatic (i.e., magnetized) electrons first need to reach a total threshold velocity along the field lines $v_g \approx (B_\perp / B_0)\,v_{\mathrm{sh}}$ which implies the geometric condition $\vec{v} \cdot \vec{B}/B > v_g$ (where $\vec{v}$ is electron velocity). If SLAMS are subluminal,\footnote{We refer here to the luminality of SLAMS and not to that of the shock, since the superluminal fields at parallel shocks are induced by SLAMS and not by the mean background field as in the case of oblique shocks.} electrons pre-heated in the precursor can fulfill this geometric condition. However, if the shock is fast (i.e., $v_{\mathrm{sh}} > c\, B_0 / B_\perp$), SLAMS present a superluminal configuration to magnetized electrons. This is the case observed in all our high-$M_{\rm A}$ PIC runs, since $B_\perp / B_0 > 1 / \gamma_{sh} \beta_{sh}$. The superluminal SLAMS are also expected for shocks with velocities $v_\mathrm{sh} \sim 10^4\,\mathrm{km/s}$ as we show in Sec.~\ref{sec:GeV}.

\begin{figure}[t!]\centering
\includegraphics[trim=0px 0px 0px 0px, clip=true,width=\columnwidth]{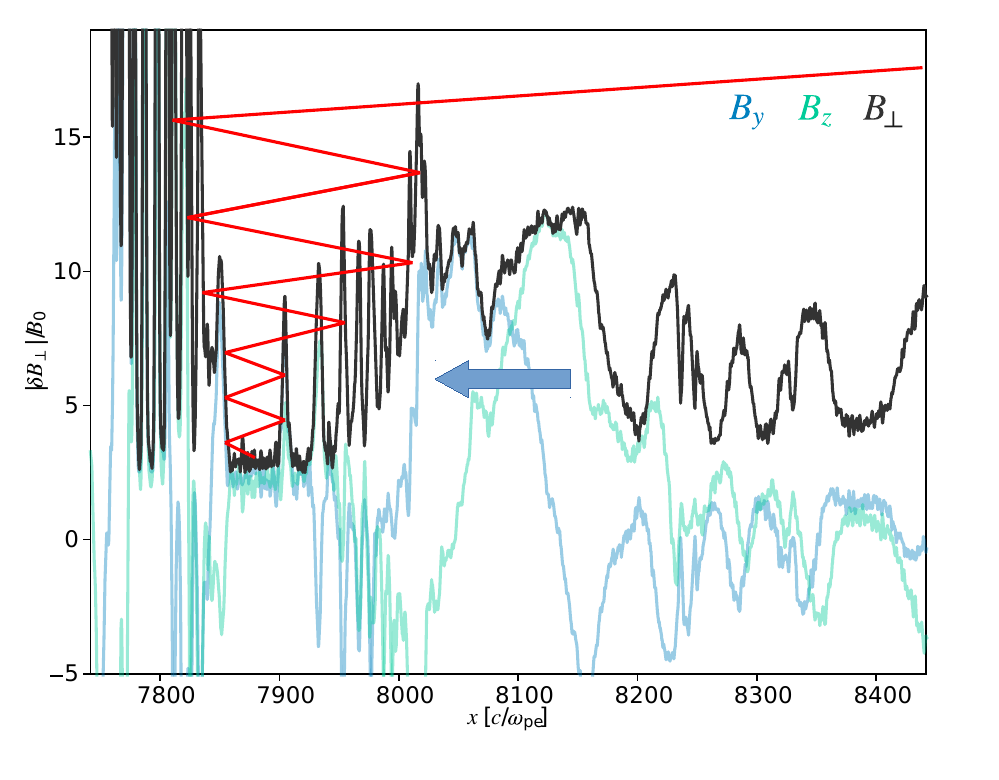}
\caption{\label{fig:qsa_scheme} 
Schematic view of electron acceleration in QSA. The plotted fields are from the late stage in Run 1: $B_\perp$ is shown in grey, $B_y$ in blue, and $B_z$ in cyan. Shock is at $x\sim 7800\,c/\omega_{pe}$, the previous SLAMS' maximum is advected behind the shock ($x < 7800\,c/\omega_{pe}$), and the peak of nearest approaching SLAM in the upstream is at $x\sim 8000\,c/\omega_{pe}$. Blue arrow marks the direction of the upstream flow (carrying SLAMS). Electron trajectory is schematically plotted in red. In the real simulation, the whole structure of SLAMS with the short scale dips (visible at $x\sim 7900-8000\,c/\omega_{pe}$) is approaching the shock, so that electron reflections become more frequent as its diffusion length shrinks.}
\end{figure}

Although at fast high-$M_{\rm A}$ shocks SLAMS impose a superluminal barrier, we also find that they quasi-periodically open a window for the peculiar Fermi I electron acceleration on SLAMS that we introduced in Sec.~\ref{sec:spectra}. Such a scenario is shown schematically in Fig.~\ref{fig:qsa_scheme} which captures the field structure in 1D Run 1, at a stage earlier than shown in Fig.~\ref{fig:1d_xl} for better clarity.
SLAMS at this stage develop a large-scale oscillating AM structure in $B_\perp$ with the largest scale $\lambda_\mathrm{SLAMS} \sim 500\,c/\omega_{pe}$ corresponding to the oscillation across the whole region $x\sim 7800 - 8300\,c/\omega_{pe}$ (similar SLAMS are also visible in the region $x\sim 5000 - 6500\,c/\omega_{pe}$ in Fig.~\ref{fig:1d_xl}f,g). The minima in such a structure form subluminal regions where magnetized electrons can almost freely propagate along the flow direction due to low inclinations of the field lines (relative to the shock normal). Inside these regions electrons thus start to gain energy by bouncing (i.e., getting mirror reflected) between the nearest approaching maximum in $B_\perp$ and the shock, as shown in electron trajectories in Fig.~\ref{fig:elec_plots} and described in the previous chapter. We observe that all magnetized electrons (within and outside the loss-cone) get mirror reflected by the maximum in the transverse field, meaning that the reflection itself is non-adiabatic. We call these Fermi I reflections \textit{quasi-periodic shock acceleration} (QSA).

Beside the subluminal regions induced by the large-scale AM field of SLAMS (e.g., oscillations on $\lambda_\mathrm{SLAMS} \sim 500\,c/\omega_{pe}$ and related shorter scales $\sim 50-100\,c/\omega_{pe}$ in the region $x > 8000\,c/\omega_{pe}$ in Fig.~\ref{fig:qsa_scheme}), the diffusion is also enhanced by electron-driven waves. Electron waves induce subluminal, short-scale dips and superluminal spikes in $B_\perp$ of the size $\sim 10 - 20\,c/\omega_{pe}$ in front of the shock in Fig.~\ref{fig:qsa_scheme} (also associated with sharp density spikes in Fig.~\ref{fig:1d_xl}g). Electron scattering on such spikes at high-$M_{\rm A}$ parallel shocks could represent a highly non-linear equivalent to the processes of SSDA (\citealt{SSDA}) and acceleration by electron-driven waves (\citealt{qperp_1d_acc}) at quasi-perpendicular shocks. Being significantly shorter than the scale of SLAMS, the dips speed up the QSA process by inducing Fermi I reflections on the spatial scales and timings comparable to that of SDA. The theory that we derive in the following chapter implies that a periodic reflective barrier should produce a power law with the same steepness at any scale (electron wave or SLAMS). However, Eq.~\ref{equ:res} implies that reflections by electron-scale spikes are constrained to the initial momenta $p_e/m c \lesssim 1$ (where electrons are magnetized with respect to the spikes), meaning that the resulting spectra is immersed in the Maxwellian in Fig.~\ref{fig:spectra}.

As the next maximum in $B_\perp$ reaches the shock, QSA diminishes and electrons there only get energized via SDA. Most of electrons that are still magnetized at this point become captured by the superluminal barrier and advect downstream, while some can still diffuse through SLAMS due to dips. The absence of accelerating magnetized electrons ahead of the shock in Figs.~[\ref{fig:1d_xl}h,\ref{fig:2d_w2800_perl}g] implies that the acceleration region is constrained to $\sim \lambda_{\rm SLAMS}$ in front of the shock. Electrons that gain enough energy in the advection time $\tau_{\rm adv} \sim \lambda_{\rm SLAMS}/v_{\mathrm{sh}}$ of a few SLAMS oscillations through the shock and reach Larmor radii in the total field corresponding to $\sim \lambda_{\rm SLAMS}/2\pi$ become unmagnetized and detach from the superluminal field to proceed with DSA. Once the maximum in $B_\perp$ is advected across the shock, QSA starts again. It means that after each $\tau_{\rm adv}$, new electrons begin their acceleration, and only those that become unmagnetized get injected into DSA. Previously, we showed that SLAMS appear on a scale comparable to the Larmor radius that corresponds to an average momentum of non-thermal ions in the ambient field $B_0$. At the same time SLAMS allow much lighter electrons to quickly reach this ion momentum by QSA and then switch to DSA. This clearly highlights the importance of SLAMS in electron acceleration up to the DSA injection energy in high-$M_{\rm A}$ astrophysical shocks.

\subsection{\label{sec:qsamod}Model of electron acceleration with SLAMS}

We now discuss a simplified model of QSA which describes the acceleration by superluminal SLAMS at fast shocks. We present the general concept here, while the detailed derivation is given in Sec.~\ref{sec:minmod} in Appendix. We assume that the magnetized, relativistic electron (with $v \approx c$) accelerating in QSA scatters from the nearest SLAMS' barrier and its shock compressed counterpart. The barrier advects with $\delta v = v_\mathrm{SLAMS} - v_\mathrm{sh}$ toward the shock, where $v_\mathrm{SLAMS}$ ($< v_\mathrm{sh}$) is its upstream speed. In the barrier rest frame, electron gets backscattered with the probability $\mathcal{P}_\mathrm{ref}$ (that depends on the properties of the barrier itself) toward the shock. In the shock frame, the ratio between the advection velocity $\delta v$ and the electron velocity $v_x$ along the shock normal then sets the probability $\mathcal{P}_\mathrm{adv} = -\delta v / v_x$ for magnetized electron to get caught and advected by the barrier. Once averaged over all possible pitch angles that backscattered electron can have, we get
\begin{equation}
    \mathcal{P}_\mathrm{adv} = \frac{\delta v}{c}\, \frac{ \delta v^2 }{ (\delta v^2 - c^2) \ln{\left( 1 + \frac{\delta v}{c} \right)} + c\,\delta v }.
\end{equation}
Assuming that all accelerating electrons get reflected from the shock, the probability $\mathcal{P}_\mathrm{QSA} = \mathcal{P}_\mathrm{ref}\cdot (1-\mathcal{P}_\mathrm{adv})$ of a particle to stay in QSA (i.e., not to get advected or transmitted through the barrier) sets the electron density $N_k$ in $k$th cycle as
\begin{equation}
    N_k = N_0\, (1-\mathcal{P})^k,\ \ \ \ln{\frac{N_k}{N_0}} = k \ln{(1-\mathcal{P})},
\end{equation}
where $N_0$ is the total number density of electrons injected into QSA, and $\mathcal{P} = 1-\mathcal{P}_\mathrm{QSA}$ is the particle probability to leave QSA.
Expression for the electron momentum in $k$th cycle (assuming the Fermi I momentum gain $\mathcal{G}$),
\begin{equation}
    \ln{\frac{p_k}{p_0}} = k \ln{(1+\mathcal{G})},
\end{equation}
is then related to the density through index $k$ as:
\begin{equation}
    \ln{\frac{N(p)}{N_0}} = \frac{\ln{(1-\mathcal{P})}}{\displaystyle \ln{(1+\mathcal{G})}}\cdot \ln{\frac{p}{p_0}},
\end{equation}
which after expanding for $\mathcal{P},\, \mathcal{G} \ll 1$ gives the momentum distribution of QSA electrons:
\begin{eqnarray}
    f(p) &&= \frac{1}{4\pi\, p^2} \frac{dN(p)}{dp} \sim p^{\displaystyle -\left( 3 + \frac{\mathcal{P}}{\mathcal{G}} \right)} \sim p^{-\delta_e},\\
    \delta_e &&= 3 + (1-\mathcal{P}_\mathrm{ref}) + \label{equ:fqsa} \\ \nonumber
    &&+\, \mathcal{P}_{\rm ref}\, \frac{3}{4}\, \frac{\delta v}{\Delta u}\, \frac{ \delta v^2 }{ (\delta v^2 - c^2) \ln{\left( 1 + \frac{\delta v}{c} \right)} + c\,\delta v }.
\end{eqnarray}
The momentum index $\delta_e$ thus has a dependence on $v_\mathrm{sh}$. In the range of typical velocities of young SNRs ($v_\mathrm{sh}\sim 0.033\,c-0.133\,c$) $\delta_e$ reaches its maximum value.
If electron reflections occur in the SLAMS and downstream frames then $u_1 = \delta v \sim 0.5-0.75\, v_{\mathrm{sh}}$ and $\Delta u = \delta v - u_2$. If we also assume $\mathcal{P}_\mathrm{ref}=0.7-0.9$ as measured for a similar barrier in \cite{Zach24}, we then obtain $f(p) \sim p^{[-4.7,-5.7]}$ for $v_\mathrm{sh} \in [0.033,0.133]\,c$ which is in the range of values observed in the case of young SNRs~\citep{bell_steepening}.

As shown in Fig.~\ref{fig:2d_w2800_perl}, SLAMS significantly speed up as they approach the shock and drive shocklets in front of the main shock. Since $v_\mathrm{SLAMS}$ becomes significant at the nearest barrier, the velocity of the upstream scattering center is $u_1 = \delta v = v_{\mathrm{sh}}-v_{\mathrm{SLAMS}}$. The slope $\delta_e \sim 5.6$ that we find in the case of 2D Run 8 (in Fig.~\ref{fig:spectra}c) then implies $v_{\mathrm{SLAMS}} = 0.5\, v_{\mathrm{sh}} \sim 0.133\,c$. This velocity is comparable to the speed at which the upstream plasma is caught by the first SLAMS' maximum at $x\sim 3100\, c/\omega_{pe}$, in front of the shock in Fig.~\ref{fig:2d_w2800_perl}f. In the case of our longest 1D Run 1 (in Fig.~\ref{fig:spectra}a) we recover the observed $\delta_e \sim 5.6$ for $v_{\mathrm{SLAMS}} = 0.53\, v_{\mathrm{sh}} \sim 0.067\,c$ which is again close to what is observed in Fig.~\ref{fig:1d_xl}. For strong initial SLAMS we get $\delta_e \sim 6$ once we assume $\mathcal{P}_\mathrm{ref}\approx 1$ which is reasonable since SLAMS do not have the dips in their structure during initial stages (i.e., initial SLAMS reflect all magnetized electrons). Our model of QSA thus describes well the slopes measured in our 1D and 2D runs during both the initial and evolved stages of SLAMS evolution.

QSA is a quasi-periodic process which, due to the strong advection induced by SLAMS, leads to a deviation of electron escape probability compared to DSA and thus produces a power law that is steeper than the DSA prediction up to the resonant energy. Due to the strong advection, QSA is not nearly as efficient as DSA, but it bridges the gap between the suprathermal and DSA injection energies. The advantage of QSA lies in its extremely high acceleration rate which comes to the fore as the maximum in $B_\perp$ approaches the shock. Electron diffusion length then shortens to that of SDA (i.e., to the Larmor radius of electron with $v\sim v_\mathrm{sh}$) in the limiting case.

\subsection{\label{sec:GeV}Application to young and fast SNRs}

The amplification of the CR-driven turbulence in the upstream depends on the shock Alfv\`enic Mach number~\citep{DamAnatoly2014}.
Our $M_\mathrm{A} < 80$ runs show that $B_\perp / B_0$ increases with $M_\mathrm{A}$ according to the resonant prediction $B/B_0 \sim \sqrt{M_{\rm A}}$ in \cite{DamAnatoly2014}. However, for $M_\mathrm{A} \gtrsim 80$ the SLAMS' amplification deviates to the non-resonant dependence $B_\perp/B_0 \sim M_{\rm A}^{3/2}$ (see \citealt{bell_inst}) and reaches $B_\perp/B_0 \gtrsim 20$ and 30 in our 1D runs with $M_{\rm A}=200$ (Run 5) and $M_{\rm A}=300$ (Run 6), respectively. In the case of young SNRs (or AGN jets), the shock velocity $\gtrsim 10^4\,\mathrm{km/s}$ with $M_{\rm A}\gtrsim 300$ thus implies $B_\perp / B_0 \gtrsim 30$. For adiabatic electrons to be able to propagate upstream of such a shock by gliding along the field lines, the field has to be subluminal (i.e., $v_\mathrm{sh}\cdot B_\perp / B_0 < c$). This condition is broken at the shocks of young SNRs where we expect the superluminal field to prevent magnetized electrons from accelerating via DSA.
The threshold for electron injection into DSA, therefore, shifts towards very high energies at which electrons become unmagnetized (with Larmor radii $\sim \lambda_{\rm SLAMS}$). For magnetized electrons, the field topology resembles to a certain extent the case of quasi-perpendicular shocks. Therefore, a different mechanism must exist that accelerates electrons to the energies observed in Fig.~\ref{fig:spectra}.

For young or fast SNRs it is very likely that QSA is the main acceleration mechanism that takes suprathermal electrons to DSA injection energies. The resulting spectra should remain steep up to the resonant energy (as shown in Fig.~\ref{fig:spectra}), then break at the resonance, and finally flatten towards higher energies to match the DSA prediction. The evolution of SLAMS observed in the long-term 1D (Fig.~\ref{fig:1d_xl}) and wide-box (Fig.~\ref{fig:2d_w2800_perl}) runs indicates a further growth of $\lambda_{\rm SLAMS}$.\footnote{Although $\lambda_{\rm CR}$ is non-resonant with CR ions, $\lambda_{\rm SLAMS}$ seems to be in a resonance with the Larmor radius defined by the average energy of CR ions. We expect $\lambda_{\rm SLAMS}$ to increase until the density of high-energy CR ions drops significantly so that it no longer affects their average energy.} During the final stage of the wide-box run, CR ions accelerate to $p_{i}/m_i c \sim 2$ and the electron spectrum remains steep at least up to $p_{e}/m_e c \sim 20$ at $T_{\rm end} \sim 204~\omega_{ci}^{-1}$. If CR ions are protons, then Eq.~\ref{equ:res} with real proton-to-electron mass ratio implies electron momentum which is much higher than $p_{e}/m_e c \sim 20$ (that we obtained for $m_i/m_e=32$).
In the case of a young SNR shock with $v_{\mathrm{sh}}\sim 10^4\,\mathrm{km/s}$ ($M_\mathrm{A}\gtrsim 300$), we get that electrons would have to reach $p_{e}/m_e c \gtrsim 1000$ (i.e., $\sim \mathrm{GeV}$ energy) with QSA to overcome the resonance and get injected into DSA. Furthermore, the QSA acceleration rate (see Sec.~\ref{sec:accrate} in Appendix) implies a time scale on the order of hours for electrons to reach GeV at such a shock. Therefore, on the timescales of the evolution of young SNRs (which are not achievable by the current computational resources), the steep non-thermal QSA tail $f(p)\sim p^{[-4.7,-5.7]}$ of CR electrons should be preserved at least up to GeV energies. This implies that the spectral index $\alpha = (\delta_e - 3)/2 \sim 0.85 - 1.35$ should be observed up to GHz band in the radio synchrotron emission of QSA electrons.

\section{\label{sec:sum}Summary and Discussion}

\noindent We summarize the most important results and implications of our study in the following points.

(i) \textit{SLAMS detection at high-Mach number shocks.} We identify short large-amplitude magnetic structures (SLAMS) as a phenomenon that characterizes the non-linear evolution of CR-driven non-resonant modes at high-Mach number parallel collisionless shocks. By propagating toward the upstream with super-Alfv\`enic velocity (in the upstream frame), SLAMS are able to significantly amplify the transverse magnetic field component $B_\perp$ and induce an amplitude modulation (AM) of the driven waves on a scale comparable to or larger than the average Larmor radius of non-thermal ions. As observed in lower Mach number configurations of the Earth bow shock (\citealt{SLAMS_Wilson}), we also find that SLAMS drive transient shocks that energize electrons via several SDA cycles.

(ii) \textit{Quasi-periodic Shock Acceleration (QSA) -- a new electron acceleration mechanism.} We find that at high-Mach number shocks the large amplification of $B_\perp$ in SLAMS significantly increases the threshold energy for electron injection into DSA. In the case studied here, which is relevant for young or fast SNRs and AGN jets with $v_{\mathrm{sh}} \gtrsim 10^4\,\mathrm{km/s}$, we find that SLAMS become superluminal. The magnetized electrons would have to glide with $v > v_{\mathrm{sh}}\cdot B_\perp / B_0 > c$ along the inclined field lines to escape the advecting SLAMS. Despite the fast SDA cycles at the shock front, most of the pre-accelerated electrons would eventually be advected downstream. However, since SLAMS induce amplitude modulation which lowers $B_\perp$ quasi-periodically and thus forms subluminal regions, a window opens for electrons to diffuse more into upstream. We find that electrons in such an environment accelerate by a new mechanism which we call \textit{quasi-periodic shock acceleration} (QSA). In QSA, electrons bounce between the shock and the nearest approaching SLAMS' maximum inside a very short region ($\lesssim \lambda_{\rm SLAMS}$) ahead of the shock. We observe that the acceleration rate of QSA is quite high and comparable to that of SDA, while the energy gain per cycle is comparable to that of DSA. However, we show that the probability of particle to get advected by the quasi-periodic superluminal magnetic barrier (imposed by SLAMS) is larger than the DSA advection probability. We analytically derive the electron QSA spectrum, $f(p) \sim p^{[-4.7,-5.7]}$, which agrees well with the electron spectra from our simulations. In the case of non-relativistic shocks, Eq.~\ref{equ:fqsa} implies that the slope of QSA electron distribution does not depend on the wavelength of SLAMS or their amplification, but only has a weak dependence on $v_{\mathrm{sh}}$. A similar spectral slope is thus expected for different Mach numbers throughout the SLAMS evolution.

(iii) \textit{Spectra expected at shocks of young or fast SNRs.} We show that SLAMS amplify the field by a factor of $\gtrsim 30$ in the case of high-Mach number shocks (e.g., Run 6 with $M_{\rm A}=300$). Since the SLAMS amplification increases with $M_{\rm A}$, a factor of $B_\perp/B_0 > 30$ is expected at shock velocities of $\sim 10^4\,\mathrm{km/s}$ ($M_{\rm A} \gtrsim 300$) which are quite typical for young SNRs. This implies the superluminal configuration for electrons, which in a case of a real electron--proton plasma significantly prolongs the QSA due to the large $\lambda_{\rm SLAMS}$ and amplification factor. The only way for magnetized (adiabatic) electrons to escape upstream is to accelerate by QSA until they reach the momentum resonant with $\lambda_{\rm SLAMS}$ (i.e., the electron motion becomes non-adiabatic once $r_{\mathrm{L},e} \sim \lambda_{\rm SLAMS}$) and thus detach from the superluminal field. As a direct consequence of QSA, we observe that the steep $f(p)\sim p^{-\delta_e} \sim p^{[-5,-5.7]}$ electron non-thermal tail extends up to the resonant energies (with $p_{{\rm res},e}/m_e c\sim 30$) at late times in our PIC runs. Translated to the real proton-to-electron mass ratio (using Eq.~\ref{equ:res}) this means that in the case of SNR shocks with $v_\mathrm{sh} \sim (1 - 4)\times 10^4$ km/s the steep QSA electron spectrum will reach the energies $E \sim 0.6 - 3.5$ GeV given the prediction of the growth of resonant waves in \cite{DamAnatoly2014a}, or $E \sim 0.6 - 16.5$ GeV if accounting for the non-resonant modes as in \cite{bell_inst}.

(iv) \textit{Comparison with other models.} Since it is a Fermi I mechanism, QSA shows great similarity with DSA. However, the core of the QSA mechanism is different from DSA. Instead of diffusion farther into the upstream, electron motion is restricted to the short, quasi-periodically shrinking region ahead of the shock. Contrary to DSA where particles gradually scatter and eventually return back to the shock, in QSA electrons move almost ballistically along the field  in subluminal regions until they get reflected (backscattered) from the maxima in the SLAMS field. The fast acceleration rate (see Eq.~\ref{equ:tqsa} in Appendix) puts the QSA timescale to be between those of SDA and DSA. Non-linear DSA (NLDSA) models of shocks with strong precursors (see~\citealt{ellison_non-linear,blasi_non-linear}) can produce a steepening of the electron spectra (due to a concave spectral shape) up to the radio spectral index $\alpha \approx 0.8$ (i.e., the momentum index $\delta_e \approx 4.6$) at relevant CR electron energies. Although such models can explain the slopes observed for historical SNRs (e.g., Tycho, Kepler, Cas A), they do not account for extragalactic SNRs and SNe with $\alpha \sim 0.8-1.2$ as demonstrated in~\cite{bell_steepening}. 

However, the model of modified NLDSA~\citep{postcursor} where ion spectra steepen due to the enhanced CR advection in the downstream, can account for the slopes that we observe in our PIC runs. Nevertheless, in our PIC runs we do not observe the Alfv\`en drift of the downstream waves or the formation of a ``postcursor" which are required for that model to produce steeper spectra. In contrast, the spectra in our QSA model steepen due to the super-Alfv\`enic drift of the upstream SLAMS. But even without this drift, the QSA mechanism still produces a steep electron power law with $\delta_e \sim 4.6-5$ for the velocities expected at young SNRs.

It was shown in~\cite{precursor_acc} that the steep spectra with slopes similar to those of QSA can also be produced by the acceleration on converging wavefronts inside the precursor itself. Although such acceleration occasionally precedes QSA in our 2D simulations, we find that electrons gain most of their energy by QSA since electron reflections become extremely frequent once the SLAMS maximum gets close to the shock.

There exists significant similarity between the electron trajectory patterns that we find in our runs (see $x-t$ and $p-t$ plots in Fig.~\ref{fig:elec_plots}) and those related to ion acceleration by oppositely propagating MHD solitons studied in a non-shock configuration by \cite{Kuramitsu_Hada_2000}. It implies that in both cases particles accelerate via Fermi I by performing non-adiabatic mirror reflections from the non-linear magnetic structures. From the point of view of a single particle, soliton acceleration represents essentially the same process as that of an electron bouncing between the shock and the nearest SLAMS maximum. However, in QSA such acceleration repeats periodically ahead of the shock, with many electrons simultaneously accelerating within each advection period of SLAMS. Since there are particle losses and electrons are mostly relativistic ($v\approx c$) the QSA spectrum appears as a power-law. The spectrum in \cite{Kuramitsu_Hada_2000} was rather bump-like since there were no particle losses and all particles were non-relativistic.

We find that the short scale dips observed in $B_\perp$-profiles in Figs.~\ref{fig:1d_xl},\ref{fig:qsa_scheme} correspond to electron-driven waves right in front of the shock. These periodically amplified waves significantly improve the QSA acceleration rate by reflecting lower-energy electrons at a distance which is much shorter than $\lambda_{\rm SLAMS}$. Electron reflections which occur on the spikes in $B_\perp$ produced by the short waves is very similar to the stochastic shock drift acceleration (SSDA; \citealt{SSDA}) at quasi-perpendicular shocks. The results of recent kinetic simulations (\citealt{Matsumoto}, \citealt{Amano}) showed that SSDA significantly contributes to electron acceleration at high Mach number quasi-perpendicular shocks. However, at parallel shocks these waves get significantly amplified and thus impose a strong, superluminal, approaching barrier to accelerating electrons. The electron mean free path in QSA is not stochastic as in SSDA. Rather, it shrinks linearly with time as electrons reflect between the shock and the approaching spikes during a single QSA cycle, which results in the steep spectra as shown in Fig.~\ref{fig:spectra}. 

In the general case of a test particle regime presented in~\citet{bell_steepening}, it is shown that higher order anisotropies can cause the steepening of the radio spectral index (of SNRs and SNe) at higher magnetic obliquities of the fast shocks with a low ratio between collision (scattering by small-scale magnetic field) and gyration frequencies $\nu / \omega_c \lesssim 0.1$. In the case of electrons accelerated by QSA, where $\nu / \omega_{ce} \sim (1/\langle\tau_\mathrm{diff}\rangle) / \omega_{ce} \sim (c/\lambda_{\rm SLAMS}) /\omega_{ce} \sim \beta_{sh}^{-1} (B_0/B_\perp) (m_e/m_i)$, we obtain $\nu / \omega_{ce} \lesssim 0.05$ for the parameters from our runs. The large amplification of $B_\perp/B_0\gtrsim10$ (i.e., the magnetic obliquity of $\gtrsim 84^\circ$) induced by SLAMS turbulence thus generates a necessary condition for steepening. In the case of superluminal SLAMS, it always leads to the corresponding radio spectral indices $\alpha \approx 0.85-1.35$ ($\delta_e \sim 4.7-5.7$). Our approach to QSA, therefore, provides insight into how a kinetic picture of electron acceleration with SLAMS at parallel shocks translates to the macroscopic model of oblique shocks  presented in~\citet{bell_steepening}.

(v) \textit{Expectations, potential caveats, and future steps.} We expect the electron spectrum to gradually flatten out (to $\sim p^{-4}$) at post-resonant energies, where the modification of the escape probability and particle mean free path become less affected by SLAMS, and QSA starts to transition to DSA. Since the electron mean free path increases significantly once DSA takes over QSA around the resonance, the electron acceleration time increases enormously. The current computational resources do not allow us to reach the end times at which a clear transition from QSA to DSA could be observed in the electron spectra. However, test particle approach may represent a useful tool bridging the gap towards realistic space and time scales. Recent test particle simulations (\citealt{Zach24}) reveal that post-resonant electrons indeed continue their acceleration via DSA. These simulations show that SLAMS not only accelerate CR electrons via QSA up to GeV energies, but are also able to accelerate electrons via DSA up to TeV energies on a time scale $\lesssim$ year.

The transverse size of our 2D run in a wide box is chosen large enough to capture the SLAMS-driven cavities and thus account for the 2D effects such as the shock corrugation, or vortex formation. For longer runs, a wider box would certainly be required to capture the cascade of cavities to the largest sizes as SLAMS continue to grow. In this case, we expect the steeper slope to be extended to correspondingly higher CR electron energies where $r_{\mathrm{L},e} \sim \lambda_{\rm SLAMS}$.

At the present stage with the end time $T_{\rm end} \sim 204~\omega_{ci}$, we also find good agreement with hybrid kinetic simulations at high Mach numbers studied in \citet{DamAnatoly2014}. The size of the SLAMS-driven cavities and field amplification and its topology in our large-box 2D run are comparable to the case of a fast shock with $M = 100$ presented in the hybrid study. This shows that it is possible to further utilize the results of hybrid or even MHD-PIC studies to probe the electron spectrum with realistic scale SLAMS.

To test our model under realistic conditions, we plan future high Mach number shock runs in 3D -- the PIC equivalent of the hybrid run in \cite{Luca23} using electron-proton plasmas and larger simulation domains.

\begin{acknowledgments}
    We thank Damiano Caprioli and Bojan Arbutina for helpful discussions. We also thank the anonymous referee for reviewing the manuscript carefully and providing constructive suggestions. This research was supported by the Multimessenger Plasma Physics Center (NSF grant PHY-2206607) and the Simons Foundation (grants 267233 and 00001470). The authors acknowledge the Texas Advanced Computing Center (TACC) for providing HPC resources Stampede2 (project TG-AST100035) and Frontera (project AST22024).
    This research also used resources of the National Energy Research Scientific Computing Center (NERSC, project ERCAP0028482) and computational resources managed and supported by Princeton Research Computing at Princeton University. VZ was financially supported by the Ministry of Education, Science and Technological Development of the Republic of Serbia through the contract No. 451-03-136/2025-03/200104.
\end{acknowledgments}

\bibliography{main}{}
\bibliographystyle{aasjournal}

\appendix

\section{\label{sec:shkinit}Shock initiation and formation of SLAMS}

In a typical shock initiation, where the plasma beam is specularly reflected off the left wall, the initially cold and dense returning ions drive transient strong instabilities in the upstream that are not representative of the developed shock environment. Since SLAMS grow from the non-resonant waves which appear later, it is crucial to wait longer for the return ions to drive those modes ahead of the shock. To speed up the shock evolution and avoid creating a strong returning beam, we use a new method where we introduce an in-plane, non-oscillating, perpendicular magnetic field component in region next to the reflecting left wall. This external field starts as $B_y = 4\,B_0$ and decreases linearly with distance from the left wall until it drops to zero after a few ion Larmor radii. The field gradient moves with the left wall (without imposing motional electric fields), and affects upstream particles that have received a kick from the piston. 
The external field with a gradient enables a gradual change of a quasi-perpendicular shock (which forms close to the left wall), into a parallel one farther in the upstream. Due to this gradual transition, reflected ions get gyrotropized and form a diffuse beam which is similar to the CR precursor that develops in the later stages of shock evolution. As a result, the strength and duration of the transient is reduced.

We find that formation of initial SLAMS proceeds as follows. When the upstream waves become significantly amplified, their wavelength becomes comparable to the Larmor radius of returning ions in the amplified field. Ions begin to slow down (which results in the appearance of ion loops in the phase space in Fig.~\ref{fig:1d_init}a) and transfer the momentum to the waves. The waves start to move in the upstream as in the case of CR current-driven instabilities studied in~\cite{MarioAnatoly2009} and, in turn, push and accelerate the upstream plasma. Once the equilibrium between the force imposed by the beam and the reaction of the upstream plasma is reached, the wave growth saturates. In the local reference frame of SLAMS, the momentum inflow from the shock side by the returning ion beam with density $n_b$ equals the momentum inflow of the background plasma with density $n_0$ from the upstream side, or $n_b\,(v_b - v_{\mathrm{SLAMS}}) = n_0\,v_{\mathrm{SLAMS}}$. The initial returning beam moves very fast ($\sim 1.5-2\,v_{{\mathrm{sh}}}$) and reaches the density $n_{\rm CR}/n_0 \sim 0.1-0.2$ (relative to the upstream plasma $n_0$) in front of the shock. SLAMS thus reach a super-Alfv\`enic velocity in the upstream frame, $v_{\mathrm{SLAMS}}\sim 2\,(n_{\rm CR}/n_0)\, v_{{\mathrm{sh}}} > v_{\rm A}$, and drive strong transient shocks in the upstream.

\section{\label{sec:mime100}SLAMS in a high mass ratio run}

\begin{figure*}[h!]
\centering
\includegraphics[trim=0px 0px 0px 0px, clip=true,width=0.45\textwidth]{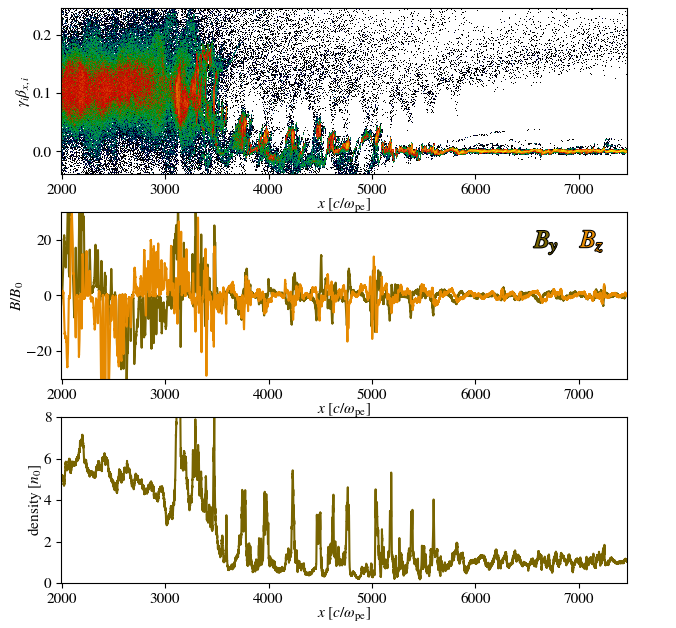}
\includegraphics[trim=0px 0px 0px 0px, clip=true,width=0.45\textwidth]{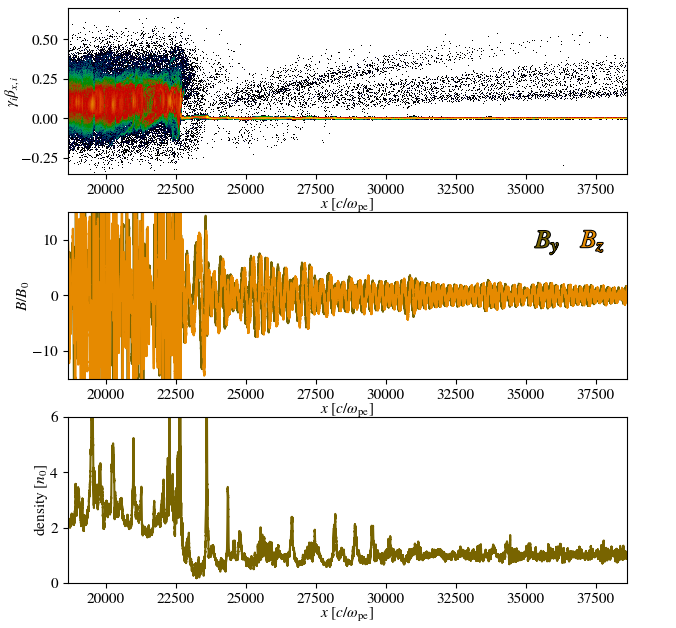}
\caption{\label{fig:1d_mime100} 
Initial SLAMS (left) and evolving SLAMS (right) in the long-term $M_{\rm A}=80$, 1D Run 4 ($m_i/m_e=100$). Ion phase space, the magnetic fields, and the density profile are shown from top to bottom (respectively) at times $t \sim 22\ \omega_{ci}^{-1}$ for the initial SLAMS and $t \sim 72\ \omega_{ci}^{-1}$ for the evolving SLAMS. The current shock is located at $x\sim 3500\ c/\omega_{pe}$ (left) and $x\sim 22500\ c/\omega_{pe}$ (right), while in both cases quasi-periodic series of shocklets are visible in the near-upstream region. In the late phase, in the region of the first few SLAMS ahead of the shock, ion gyrations can be observed.}
\end{figure*}

We present the initial and evolving SLAMS in 1D Run 4 which has the higher ion-to-electron mass ratio $m_i / m_e = 100$ (see Fig.~\ref{fig:1d_mime100}). The SLAMS appear with a similar amplification of $\sim 10$ and evolve with a similar spatial scale compared to the fiducial run with $m_i / m_e = 32$ (Run 1). This is to be expected since we use the same values $v_{\mathrm{sh}}=0.133~c$ and $M_{\rm A}=80$ in both cases, so that the changes in $m_i/m_e$ and $B_0$ then lead to comparable growth rates and wavelengths of the CR-driven (Bell) modes, which is also true for the initial SLAMS. However, the evolution of SLAMS is slower in later phases at higher mass ratio since $\lambda_{\rm SLAMS}$ is defined by the Larmor scale of CR ions. Only at the end of the run do electrons begin to form a steep power law. The faster evolution of SLAMS (and thus particle acceleration) makes $m_i / m_e = 32$ our preferred choice for most runs.

\section{\label{sec:minmod}Derivation of the minimal model of electron acceleration with SLAMS}

In our simplified model of QSA, which is best suited to superluminal SLAMS at fast shocks, we assume that the energy gain in each cycle is comparable to that of DSA. We also assume that the diffusion length $\lambda_{\rm diff}$ shrinks from $\lambda_{\rm SLAMS}$ to $r_{\mathrm{L},e}$ over the advection time $\tau_{\rm adv}$ of each SLAMS' oscillation as:
\begin{equation}
    \lambda_{\rm diff} \sim \lambda_{\rm SLAMS} \left( 1 - \frac{t \bmod \tau_{\rm adv}}{\tau_{\rm adv}} \right).
\end{equation}
We set $B_\perp \sim 0$ within the subluminal regions (minima in $B_\perp$). The pre-heated electrons ($p \sim m_e c$) that are injected (into QSA) at the shock thus freely move across the subluminal region in front of the shock. The electrons encounter a superluminal SLAMS' barrier with amplified $B_\perp$ farther in the upstream. We set this barrier to be partly permeable so that it reflects and isotropizes impinging electrons with the constant probability $\mathcal{P_\mathrm{ref}}$ at each encounter. The back-scattered electrons then travel to the shock where they experience another Fermi I reflection (toward the upstream) and enter the next QSA cycle. The derivation that follows will thus be valid for any cycle (it repeats with each cycle).

We start with the generalized form assuming that a pre-heated, magnetized electron has already began bouncing between the shock and the approaching SLAMS' maximum (i.e., barrier). In the $n$th cycle of acceleration, the electron starts at the leading shock edge (e.g., being reflected from the shock) and glides upstream along the parallel background field lines $B_0$ toward the nearest approaching barrier as shown in Fig.~\ref{fig:qsa_scheme}. We set the positive $x$-axis to point toward the upstream (as it appears in all figures). In the shock frame, the barrier then moves with the speed $-\delta v = -(v_\mathrm{sh} - v_\mathrm{SLAMS})$, where $v_\mathrm{sh}$ and $v_\mathrm{SLAMS}$ are the shock and SLAMS' velocities in the upstream frame, respectively. In the barrier rest frame, the electron gets reflected and scattered to some arbitrary direction due to the isotropization. The electron longitudinal velocity after the reflection in the barrier (primed) frame is $v_x' = \mu v$, where $\mu = \cos{\theta} \in [-1,0]$ and $\theta$ is the pitch angle. In the shock frame this velocity transforms to
$$v_x = \frac{\mu v - \delta v}{1 - \displaystyle\frac{\mu v\, \delta v}{c^2}}.$$
Both the electron and the barrier then move toward the shock with the velocities $v_x$ and $-\delta v$, respectively.
We define the probability that the back-scattered electron (with $v \to c$) is caught by the barrier as a ratio between these velocities:
\begin{equation}
    \mathcal{P}_{\rm adv} = \frac{-\delta v}{v_x} = \frac{-\delta v}{c}\, \frac{c - \mu \delta v}{\mu c - \delta v}.
\end{equation}
If electron is reflected parallel to the direction of the upstream flow ($\mu = -1$) the probability of electron to be caught and advected by the barrier is small ($\delta v / c$). If electron is scattered perpendicular to the flow, then $\mathcal{P}_{\rm adv} = 1$ and electron gets caught and trapped by the barrier. Since electrons are scattered to an arbitrary negative $x$-direction in the barrier frame, we use the averaged velocity flux of all backscattered electrons:
\begin{equation}
    \langle \mathcal{P}_{\rm adv} \rangle = \frac{-\delta v}{c}\, \left( \int_{-1}^{0} \frac{\mu c - \delta v}{c - \mu \delta v} \, d\mu \right)^{-1} = \frac{\delta v}{c}\, \frac{ \delta v^2 }{ (\delta v^2 - c^2) \ln{\left( 1 + \frac{\delta v}{c} \right)} + c\,\delta v }.\label{equ:probadv}
\end{equation}
The probability that the relativistic, magnetized electron is reflected and not caught by the advecting barrier or transmitted (i.e., it remains in QSA) is then:
\begin{equation}
    \mathcal{P}_{\rm QSA} = \mathcal{P}_{\rm ref} \cdot (1 - \langle \mathcal{P}_{\rm adv} \rangle) = 1 - (1-\mathcal{P}_{\rm ref}+\mathcal{P}_{\rm ref}\cdot\langle\mathcal{P}_{\rm adv}\rangle) = 1 - \mathcal{P}.\label{equ:probqsa}
\end{equation}

Knowing the QSA probability, we use the same method as in \citet{kappa} to derive the distribution of QSA electrons. The equations for the cumulative density change and momentum gain in the QSA case are (respectively):
\begin{eqnarray}
    \frac{d N}{N} &&= d \ln{N} = -\mathcal{P} = \mathcal{P}_{\rm QSA} - 1 = - (1-\mathcal{P}_{\rm ref}) - \mathcal{P}_{\rm ref}\cdot \frac{\delta v}{c}\, \frac{ \delta v^2 }{ (\delta v^2 - c^2) \ln{\left( 1 + \frac{\delta v}{c} \right)} + c\,\delta v },\label{equ:dn} \\
    \frac{d p}{p} &&= d \ln{p} = \mathcal{G} = \frac{4}{3}\, \frac{\Delta u}{v},\label{equ:dp}
\end{eqnarray}
where $\mathcal{P}$ is the electron escape probability; $\mathcal{G}$ is the momentum gain; $v \approx c$ is the velocity of pre-energized adiabatic electrons; and $\Delta u$ is the differential velocity between the reference frames of electron scattering centers (i.e., the approaching SLAMS' maximum and its shock-compressed predecessor). For the cumulative distribution $dN = - 4\pi p^2 f dp$ the previous equations imply:
$$
    \frac{d \ln{\left(\frac{\mathcal{G}}{\mathcal{P}} f \right)}}{d \ln{p}} = -\left( \frac{\mathcal{P}}{\mathcal{G}} + 3 \right).
$$
After integration, we obtain a power-law distribution for the QSA mechanism:
\begin{eqnarray}
    &&\ln{f(p)} \sim -\delta_e \ln{p},\\ \nonumber
    &&\delta_e = 3 + (1-\mathcal{P}_{\rm ref}) + \mathcal{P}_{\rm ref}\cdot \frac{3}{4}\, \frac{\delta v}{\Delta u}\, \frac{ \delta v^2 }{ (\delta v^2 - c^2) \ln{\left( 1 + \frac{\delta v}{c} \right)} + c\,\delta v }.
\end{eqnarray}

\subsection{\label{sec:accrate}Acceleration rate}

At the moment electron departs from the shock with the velocity along the shock normal $v_x = \mu_{+} c$, the barrier is at the distance $d_k$ farther ahead. In the shock frame, the electron and the barrier meet at the distance $\mu_{+}\, c\cdot \delta t_{+}$ ahead of the shock. During the same interval $\delta t_{+}$, the barrier gets advected by the distance $(v_{\mathrm{sh}}-v_{\mathrm{SLAMS}})\, \delta t_{+} = \delta v\, \delta t_{+}$ toward the shock, which means that $d_k = (\mu_{+}\, c + \delta v)\,\delta t_{+}$. At the meeting point, the electron gets reflected and scattered to $\mu_{-}$ in the barrier rest frame. The electron then travels the same distance back to the shock over the time $\delta t_{-} \approx \mu_{+}\,c / (\mu_{-}\,c - \delta v) \cdot \delta t_{+}$ to reach the shock again. The time for the electron to complete the $k$th cycle in QSA is then $t_k = \delta t_{+} + \delta t_{-} = \delta t_{+} (1 + \mu_{+}\,c / (\mu_{-}\,c - \delta v))$. Averaging over $\mu_{+} \in [0,1]$ and $\mu_{-} \in [-1,0]$ gives:

$$\langle t_k \rangle = \frac{d_k}{c}\, \frac{\delta v}{c}\, \ln^2{\left( 1 + \frac{c}{\delta v} \right)} = \frac{\delta v}{c^2}\,\ln^2{\left( 1 + \frac{c}{\delta v} \right)}\, \left( d_0 - v_{\mathrm{sh}} \sum_{i=0}^{k-1}t_i \right),$$ where $0<d_0<\lambda_\mathrm{SLAMS}$ is a distance of the SLAMS' maximum from the shock, at the time when electron starts its zeroth cycle. We further drop the bracket notation in $\langle t_k \rangle$ and use $t_k$ instead. From the previous averaged relation we get:
\begin{eqnarray}
    t_k &=& t_{k-1}\left[ 1 - \frac{\delta v^2}{c^2}\, \ln^2{\left( 1 + \frac{c}{\delta v} \right)} \right] = \nonumber \\
    &=& t_0 \left[ 1 - \frac{\delta v^2}{c^2}\, \ln^2{\left( 1 + \frac{c}{\delta v} \right)} \right]^{k-1},\,t_0 = \frac{d_0}{c}\, \frac{\delta v}{c}\, \ln^2{\left( 1 + \frac{c}{\delta v} \right)}. \nonumber
\end{eqnarray}

The total QSA acceleration time (to complete $k$ cycles) is then:

$$\tau_\mathrm{QSA} = \sum_{i=0}^{k} t_i = \frac{d_0}{\delta v} \left[ 1 - \left( 1 - \frac{\delta v^2}{c^2}\, \ln^2{\left( 1 + \frac{c}{\delta v} \right)} \right)^k \right].$$

We switch to the discrete form of momentum gain in Eq.~\ref{equ:dp} by substituting $dp/p \to p_{k+1}/p_k - 1$ to get $p_k/p_0 = (1+\mathcal{G})^k$ (where $p_0/mc \sim 1$ is electron injection momentum). Then we can relate the previous expression directly to the momentum $p(k) \equiv p_k$ in a continuous form through index $k = \ln{(p/p_0)} / \ln{(1+\mathcal{G})}$ as:

$$\tau_\mathrm{QSA} = \frac{d_0}{\delta v} \left[ 1 - \left( 1 - \frac{\delta v^2}{c^2}\, \ln^2{\left( 1 + \frac{c}{\delta v} \right)} \right)^{\displaystyle\frac{\ln p/p_0 }{\ln (1+\mathcal{G}) }} \right],$$
which, in the limit of non-relativistic shocks and after averaging over a single advection time $\tau_{\rm adv}=\lambda_\mathrm{SLAMS}/\delta v$, reduces to:
\begin{equation}
    \tau_\mathrm{QSA} \sim \tau_\mathrm{adv} \left[ 1 - \left(\frac{p}{p_0} \right)^{-\displaystyle \frac{\delta v}{c}\, \ln^2{\left( 1 + \frac{c}{\delta v} \right)}} \right].
    \label{equ:tqsa}
\end{equation}
The QSA acceleration time is therefore comparable to the advection time of a single SLAMS' oscillation for most electron momenta.

Electrons that initially enter SDA are pre-energized inside the precursor and reach the shock with a significant momentum $p_0/m_e c \sim 1$ (i.e., the Larmor radius $\sim p_0/p_{sh} \cdot M_\mathrm{A} \sqrt{m_i/m_e}^{-1}\,c/\omega_{pe}\sim 100\,c/\omega_{pe}$). Due to such a large Larmor radius, electrons are able to sample the full velocity gradient at the shock (implying the momentum gain $\sim \mathcal{G}$) at each SDA gyration. Using the discrete form of Eq.~\ref{equ:dp} as in the previous case of QSA, we can estimate the SDA acceleration time as a time it takes for electron to reach the momentum $p$ after $k$ cycles of SDA:
\begin{equation}
    \tau_\mathrm{SDA} \approx \frac{\pi}{\omega_{ce}}\cdot k \approx \frac{\pi}{\omega_{ce}}\cdot \beta_{sh}^{-1} \ln{\frac{p}{p_0}}.
    \label{equ:tsda}
\end{equation}

\end{document}